%% file: paper.tex
\documentclass[11pt,a4paper]{article}
\usepackage[colorlinks=true, linkcolor=black!50!blue, urlcolor=blue, citecolor=blue, anchorcolor=blue]{hyperref}
\usepackage[font=small,labelfont=bf,margin=0mm,labelsep=period,tableposition=top]{caption}
\usepackage[a4paper,top=1cm,bottom=1.5cm,left=1.5cm,right=1.5cm,bindingoffset=0mm]{geometry}

\usepackage{graphicx}
\usepackage{float}
\usepackage{afterpage}
\usepackage{epsfig,cite}
\usepackage{amssymb}
\usepackage{amsmath}
\usepackage{dsfont}
\usepackage{multirow}
\usepackage{url,hyperref}
\usepackage{booktabs}
\usepackage{tabularx}
\usepackage{xcolor}
\usepackage{bm}
\usepackage[symbol]{footmisc}
\usepackage{textcomp}
\usepackage{caption}
\usepackage{amsmath,graphicx,slashed,color}
\usepackage{graphicx,epsfig,amssymb,bm}
\usepackage[utf8]{inputenc}
\usepackage{physics}
\usepackage{mathtools}
\usepackage{todonotes}

\usepackage{algorithm2e}
\usepackage{optidef}

\usepackage{tabularx}
\newcolumntype{C}[1]{>{\centering\arraybackslash}p{#1}}

\usepackage[normalem]{ulem}

\begin{document}

\vspace{-2.0cm}

\begin{center} 

  {\Large \bf Regularising experimental correlations in LHC data:\\
    \vspace{0.2cm}
    theory and application to a global analysis of parton distributions
  }
  \vspace{.7cm}

  Zahari Kassabov$^1$, Emanuele R. Nocera$^2$, Michael Wilson$^2$

\vspace{.3cm}
{\it ~$^1$ DAMTP, University of Cambridge, Wilberforce Road, Cambridge CB3 0WA, United Kingdom}\\
{\it ~$^2$ The Higgs Centre for Theoretical Physics, University of Edinburgh,\\
JCMB, KB, Mayfield Rd, Edinburgh EH9 3JZ, United Kingdom\\}

\end{center}

\vspace{0.1cm}

\begin{center}
  {\bf \large Abstract}\\
\end{center}

We show how an inaccurate determination of experimental uncertainty
correlations in high-precision LHC measurements may undermine the reliability
of the associated $\chi^2$. We formulate the problem rigorously, and devise
a regularisation procedure that increases the stability of the $\chi^2$ by
altering the covariance matrix of the measurement as little as possible. We
apply the procedure to the NNPDF4.0 global analysis of parton distribution
functions that utilises a large amount of LHC measurements. We find that the
regularised $\chi^2$ of the NNPDF4.0 determination is lowered by about
$3\sigma$, without significantly altering the resulting PDFs upon refitting.

\input{sec-introduction}
\input{sec-stability}
\input{sec-regularisation}
\input{sec-PDFs}
\input{sec-conclusions}
\input{sec-acknowledgements}
\appendix
\input{app-glossary}
\input{app-boundproof}

\bibliography{paper}

\end{document}

%% file: sec-introduction.tex
\section{Introduction}
\label{sec:introduction}

With the restart of the Large Hadron Collider (LHC), particle physics is getting
into the thick of a new era, whereby measurements are anticipated to attain an
unprecedented, percent-level, statistical precision~\cite{Mangano:2020icy}.
These measurements are utilised to improve the determination of Standard
Model (SM) parameters~\cite{Azzi:2019yne}, to constrain Parton Distribution
Functions (PDFs)~\cite{Ethier:2020way}, to evaluate backgrounds for missing
energy searches~\cite{CidVidal:2018eel}, or more generally to constrain
higher-dimensional operators in the SM Lagrangian~\cite{Brivio:2017vri}, and
eventually to stress-test the properties of the Higgs
boson~\cite{Cepeda:2019klc}.

In all of these cases, measurements are contrasted with theoretical predictions
by means of statistical inference: a model is chosen and compared or optimised
to the data through maximum likelihood estimation. The goal is to reject a test
hypothesis or to obtain a confidence interval of model parameters. Because
experimental uncertainties in LHC measurements are commonly assumed to be
Gaussian, the figure of merit utilised for the statistical test or to optimise
the model is the $\chi^2$ statistic, which is monotonic in the likelihood of
sampling the experimental data given the theory.

The robustness of the $\chi^2$ as a figure of merit relies on the accuracy of
theoretical expectations and of experimental uncertainties. In this paper we
assume that the $\chi^2$ is not spoiled by inaccuracies in theoretical
expectations --- a fact that is possibly not true now, but that will become
increasingly realistic in the future~\cite{Heinrich:2020ybq} --- and focus only
on inaccuracies of experimental uncertainties.

A proper estimation of uncertainties in LHC measurements is indeed becoming
increasingly delicate. The large event samples collected during Run I and II 
have been making statistical uncertainties generally smaller than systematic
uncertainties; the upcoming Run III will make the former even smaller.
In contrast to statistical uncertainties, systematic uncertainties (which are
not related to event counts but, {\it e.g.}, to limitations of the detector or
to assumptions made in their modelling) are more difficult to estimate. The
reason being that custom procedures and subjective choices are involved in
their quantification~\cite{Sinevo}. Furthermore, systematic uncertainties are
usually correlated across different kinematic bins, both within the same
measurement and across different measurements. Determining these correlations
is even more difficult, and often it is not even attempted. In this case, simple
assumptions, such as taking systematic uncertainties to be fully correlated
or fully uncorrelated, may misrepresent the truth. More elaborated guesswork
can be performed in order to devise {\it ad-hoc} correlation models, which
however have no generality and can be time consuming.

Because the uncertainties on LHC measurements are being increasingly dominated
by systematic uncertainties, any analysis that utilises them is implicitly
dependendent on the choices made in their characterisation. While this
dependence is generally unavoidable, some care must be taken to prevent it from
hampering the use of the data in precision physics analyses based on statistical
inference.

The aim of this paper is to formulate and address this problem rigorously. We
first demonstrate how inaccuracies in the estimation of systematic uncertainty
correlations, even if small, can lead to instabilities in the experimental
covariance matrix and how these can ultimately undermine the reliability of the
$\chi^2$. We then devise a regularisation procedure whereby these instabilities
are removed with minimal information on their source, and without loss of
generality. The idea is to define a bound on the singular values of the
correlated part of the matrix of uncertainties, and to clip them to a suitably
chosen value that alters only the small subset of directions associated to
instability. We finally apply this procedure to a particular problem relevant to
LHC precision physics that utilises statistical inference: PDF determination.
Although we orient our discussion towards this problem, our regularisation
procedure is completely general, and can be applied to any statistical analysis
that involves the evaluation of the $\chi^2$.

The structure of the paper is as follows. In Sect.~\ref{sec:stability} we
introduce the matrix of uncertainties and formulate a stability criterion for
it. In Sect.~\ref{sec:regularisation} we derive our regularisation procedure
and demonstrate how it works with a toy model. In Sect.~\ref{sec:PDFs} we apply
the procedure to PDF determination using the recently released NNPDF4.0 parton
sets~\cite{NNPDF:2021njg}. We summarise our results in
Sect.~\ref{sec:conclusions}. The paper is completed by two appendices:
Appendix~\ref{app:glossary} is a glossary of some useful definitions
used through the paper; Appendix~\ref{app:boundproof} contains the proof of
Eq.~\eqref{eq:norminequality}. Our regularisation procedure is made publicly
available as part of the {\sc NNPDF} software~\cite{NNPDF:2021uiq}.

%% file: sec-stability.tex
\section{Matrix of uncertainties and its stability}
\label{sec:stability}

In this section we formulate the problem of the reliability of the $\chi^2$ if
instabilities, even if small, appear in the covariance matrix that enters its
computation. We first introduce the matrix of uncertainties and write the
$\chi^2$ in terms of it. We then derive  an upper bound on the instability of
the matrix of uncertainties that ensures the stability of the $\chi^2$ with
minimal information.

\subsection{Matrix of uncertainties and $\chi^2$}
\label{subsec:matrix_of_uncertainties}

The format of LHC measurements, as often made public through the {\sc HepData}
repository~\cite{Maguire:2017ypu}, consists of a central value and of a set of
uncertainties for each of the data points that form the measurement itself. The
set of uncertainties usually encompass a total statistical uncertainty and a
set of independent systematic uncertainties. The latter are typically correlated
across data points, by an amount that may be specified or not. 

Let us consider an experimental measurement made of $N_{\rm dat}$ data points,
each of which has $N_{\rm err}$ independent uncertainties. We call $\vb{d}$
the vector of experimental mean values, $\vb{d}=\{D_i\}$, and $A$ the
$N_{\rm dat}\times N_{\rm err}$ matrix of uncertainties, $A=\{A_{ij}\}$, with
$i=1, \dots, N_{\rm dat}$, and $j=1, \dots, N_{\rm err}$. Assuming that all
uncertainties are Gaussian and that they are combined additively, the
experimental measurement defines a multi-Gaussian distribution with mean
$\vb{d}$, given by the experimental central values, and covariance matrix
$C$, given by the product of the matrix of uncertainties and its transpose,
$C=AA^t$. 

Depending on the information provided with a given experimental measurement,
each element of the matrix of uncertainties can be obtained from knowledge of
$C$, for example by taking its Cholesky decomposition, or from direct knowledge
of experimental uncertainties. In this latter case, should $\mathcal{O}_i$ be
the physical observable corresponding to the data points $D_i$, and $\{u_j\}$
the set of independent variables which contribute to the experimental
uncertainty and on which the observable depends (each described by a Gaussian
distribution with central value $u_j^0$ and uncertainty $s_j$), any element of
the matrix of uncertainties reads as
\begin{equation}
  A_{ij}
  =
  \left.\frac{\partial\mathcal{O}_i}{\partial u_j} \right|_{\vb{u}=\vb{u}^0}
  s_j\,.
  \label{eq:error_matrix}
\end{equation}
If a given source of uncertainty $u_l$ affects a single data point $k$, then
$\partial\mathcal{O}_i/\partial u_l=0$ for $i\neq k$, and it corresponds to
a row in $A$ with a single non-zero entry $A_{kl}$. For instance, this is the
case for statistical uncertainties that originate from bin-by-bin event counts.
These uncertainties, together with similarly fully uncorrelated systematic
uncertainties, can therefore be encoded in a $N_{\rm dat}\times N_{\rm dat}$
diagonal sub-matrix of $A$. We assume that such uncertainties are always
present in a measurement, therefore we will henceforth consider that
$N_{\rm err} \geq N_{\rm dat}$, and that both $A$ and $C$ be full rank.

The inverse of the covariance matrix $C$ is $C^{-1}=A^{+t}A^+$, where $A^+$ is the
right inverse of $A$ (see Appendix~\ref{app:glossary}). Denoting with
$\vb{t}=\{T_i\}$, $i=1,\dots,N_{\rm dat}$, the vector of theoretical predictions
corresponding to the data mean values $\vb{d}$, the $\chi^2$ can be written as
\begin{equation}
  \chi^2
  =
  (\vb{d} - \vb{t})^t
  C^{-1}
  (\vb{d} - \vb{t})
  =
  \norm{A^+(\vb{d}-\vb{t})}^2\,.
  \label{eq:chi2}
\end{equation}
In this equation we have explicitly factorised the two contributions that
determine the value of the $\chi^2$: the difference between the mean
experimental central values and the theoretical expectation values, encoded in
$\vb{d}-\vb{t}$; and the experimental uncertainties, encoded in the error
matrix $A$.

Concerning the $(\vb{d} - \vb{t})$ term in Eq.~\eqref{eq:chi2}, we assume
perfect knowledge of theoretical expectations. This means that the vector of
differences $\vb{d}-\vb{t}$ is a realisation of a random variable which follows
a multivariate Gaussian distribution with mean zero and covariance matrix $C$.
The corresponding probability density can be given in terms of the matrix of
uncertainties $A$ and of a vector of $N_{\rm err}$ independent standard Gaussian
random variables, $\vb{n}=\left\{n_j \right\}$, $j=1,\dots,N_{\rm err}$,
\begin{equation}
  \vb{d}-\vb{t} = A\vb{n}\,,
  \qquad
  \vb{n}\sim\mathcal{N}(\vb{0},I)\,.
  \label{eq:multiGaussian}
\end{equation}

Concerning the matrix of uncertainties $A$ in Eq.~\eqref{eq:chi2}, we consider
two different cases. The first case corresponds to assuming that $A$ has
been estimated accurately. Substituting Eq.~\eqref{eq:multiGaussian} in
Eq.~\eqref{eq:chi2}, we obtain that the expected value of the $\chi^2$ over
samples of $\vb{d}-\vb{t}$ is
\begin{equation}
  \expval{\chi^2} = \expval{\norm{A^+A\vb{n}}^2}
  \,.
  \label{eq:chi2_expt}
\end{equation}
Using the fact that, for independent standard Gaussian variables,
$\expval{n_j n_l} = \delta_{jl}$, we find that $\expval{\chi^2}$ is given in
terms of the Frobenius norm (see Appendix~\ref{app:glossary}) of $A^+A$:
\begin{equation}
  \expval{\chi^2} = \sum_{j,l}^{N_{\rm err}} (A^+A)_{j,l}^2 = \norm{A^+A}_F^2
  = N_{\rm dat}
  \,,
  \label{eq:chi2asfrob}
\end{equation}
where the last equality follows from the singular value decomposition of $A^+$,
see Appendix~\ref{app:glossary}.

The second case corresponds to assuming that there are inaccuracies in the
estimation of uncertainties, which do not need to be large. We define as
$\bar{A}$ the matrix of uncertainties that contains such inaccuracies. This is
different from $A$, which is therefore unknown. The covariance matrix used to
compute the $\chi^2$ is now $\bar{C}=\bar{A}\bar{A}^t$. However, because we
still assume perfect knowledge of theoretical expectations,
Eq.~\eqref{eq:multiGaussian} continues to hold. Therefore, in analogy with
Eqs.~\eqref{eq:chi2_expt}-\eqref{eq:chi2asfrob}, the expectation value of the
$\chi^2$ reads as
\begin{equation}    
  \expval{\bar{\chi}^2} = \norm{\bar{A}^+ A}_F^2
  \,.
  \label{eq:badchi2asfrob}  
\end{equation}
A comparison between Eq.~\eqref{eq:badchi2asfrob} and Eq.~\eqref{eq:chi2asfrob}
allows one to formulate a stability criterion for the expectation value of the
$\chi^2$ and for the matrix of uncertainties $A$ upon substituting $A$ with
$\bar{A}$, as we explain next.

\subsection{Stability criterion}
\label{subsec:stability_criterion}

We state that the matrix of uncertainties $A$ is stable upon the replacement $A
\to \bar{A}$ in the computation of the $\chi^2$ when the difference in its
expected value, $\Delta\chi^2$, is smaller than statistical fluctuations of the
$\chi^2$ statistic itself, as measured by the standard deviation of the
corresponding $\chi^2$ distribution, which is equal to $\sqrt{2N_{\rm dat}}$.
We can therefore write a stability criterion for the expectation value of the
$\chi^2$ as:
\begin{equation}
  \Delta\chi^2
  =
  \abs{\expval{\bar{\chi}^2} - \expval{\chi^2}}  < \sqrt{2N_{\rm dat}}   
  \,.
  \label{eq:stabilitysimple}  
\end{equation}
Substituting Eqs.~\eqref{eq:chi2asfrob} and~\eqref{eq:badchi2asfrob} in
Eq.~\eqref{eq:stabilitysimple}, we can equivalently write
\begin{equation}
    \norm{\bar{A}^+A}_F^2 - N_{\rm dat} < \sqrt{2N_{\rm dat}} 
    \,.
    \label{eq:deltachi2fromabar}
\end{equation}

We now seek to find an upper bound to the inaccuracies of the matrix $\bar{A}$
that satisfies the stability criterion on the $\chi^2$,
Eq.~\eqref{eq:deltachi2fromabar}, using minimal information. To this purpose,
we write the matrix of uncertainties $A$, which we do not know, as a
perturbation to the matrix $\bar{A}$, which we are given,
\begin{equation}
  A = \bar{A} + \delta F    
  \,,
  \label{eq:abarfromF}
\end{equation}
where $F$ is a matrix of perturbations and $\delta$ is a scalar parameter
controlling the size of the fluctuation. We assume that $\delta$ is
sufficiently small that we can linearly expand around
$\delta=0$. Replacing Eq.~\eqref{eq:abarfromF} into
Eq.~\eqref{eq:badchi2asfrob}, we find
\begin{equation}
  \expval{\bar{\chi}^2} = \norm{ \bar{A}^+(\bar{A} + \delta F) } _F^2
  \,.
\end{equation}
Using the triangle inequality, we can derive the upper bound
\begin{equation}
  \expval{\bar{\chi}^2}
  \leq
  \left(\norm{\bar{A}^+\bar{A}}_F  + \delta \norm{\bar{A}^+F}_F \right)^2
  \,.
\end{equation}
Then expanding the square, and using the fact that $\bar{A}$ is full rank since
it corresponds to the covariance matrix obtained in the experimental analysis,
we obtain
\begin{equation}
  \expval{\bar{\chi}^2}  \leq  N_{\rm dat}
  + 2 \delta \sqrt{N_{\rm dat}}  \norm{\bar{A}^+F}_F
  + \order{\delta^2}
  \,.
  \label{eq:deltachi2bound0}
\end{equation}
Now, neglecting the quadratic terms in $\delta$, we arrive at
\begin{equation}
  \Delta \chi^2 \leq 2 \delta \sqrt{N_{\rm dat}} \norm{\bar{A}^+F}_F
  \,.
  \label{eq:deltachi2bound1}
\end{equation}
We apply the following inequality 
\begin{equation}
  \norm{XY}_F \leq \min\left(\norm{X}_F\norm{Y}_2, \norm{X}_2\norm{Y}_F\right)
  \,,
  \label{eq:norminequality}
\end{equation}
which holds for arbitrary matrices $X$ and $Y$ of compatible shape (see
Appendix~\ref{app:boundproof} for a proof), to
Eq.~\eqref{eq:deltachi2bound1} and find
\begin{equation}
  \Delta \chi^2 \leq  2 \delta \sqrt{N_{\rm dat}} \norm{\bar{A}^+}_2\norm{F}_F
  \,,
  \label{eq:deltachi2fromF} 
\end{equation}
where $\norm{\bar{A}^+}_2$ denotes the Euclidean (or $L^2$) norm of $\bar{A}$
(see Appendix~\ref{app:glossary}). Choosing $\norm{\bar{A}^+}_2\norm{F}_F$
instead of $\norm{\bar{A}^+}_F\norm{F}_2$ as bound in
Eq.~\eqref{eq:deltachi2fromF} results in tighter constraints. The reason being
that, in practice, instabilities occur when $\bar{A}^+$ has large singular
values. 

Finally, combining Eq.~\eqref{eq:deltachi2fromF} with
Eq.~\eqref{eq:stabilitysimple}, we conclude that the condition
\begin{equation}
  \norm{\bar{A}^+}_2\norm{F}_F \leq \frac{1}{\sqrt{2}\delta}
  \label{eq:finalstability}
\end{equation}
is sufficient to avoid that the expectation value of the $\chi^2$ overestimates
its true value by an amount larger than its statistical fluctuation. The
advantage of Eq.~\eqref{eq:finalstability} with respect to
Eq.~\eqref{eq:deltachi2fromabar} is to provide a stability criterion
that does not depend on the unknown matrix of uncertainties $A$, but only on
the Frobenius norm of the matrix of fluctuations $F$. This dependence can be
easily modelled as we explain in the next section.

%% file: sec-regularisation.tex
\section{Regularising the matrix of uncertainties}
\label{sec:regularisation}

In this section we devise a procedure to regularise the matrix of uncertainties
in such a way that the $\chi^2$ becomes insensitive to inaccuracies in the
estimation of the experimental uncertainties. We then demonstrate the
effectiveness of the procedure in a toy model that is representative of
realistic LHC measurements.

\subsection{Regularisation procedure}
\label{subsec:regularisation_procedure}

Our aim is to obtain a regularised matrix of uncertainties $A_{\rm reg}$ which,
for a given model of instabilities, fulfills the following criteria: {\it i})
$A_{\rm reg}$ is more stable that $\bar{A}$; {\it ii}) $A_{\rm reg}$ is compatible
with $\bar{A}$ within the precision with which this is determined; and
{\it iii}) the uncertainty estimated by $A_{\rm reg}$ never decreases in
comparison to that estimated by $\bar{A}$. To this purpose, we first need to
characterise the inaccuracies in the matrix $\bar{A}$, by means of a simplified
model that builds upon the stability criterion, Eq.~\eqref{eq:finalstability}.
We note that sometimes such a characterisation comes as part of the measurement
itself, generally as the result of a dedicated analysis. In these cases, this
characterisation has to be preferred to the model discussed below.

The model of inaccuracies that we devise ought to be minimal, general, and
realistic. Minimal, because it should alter the matrix of uncertainties
$\bar{A}$ as little as possible; general, because it should be applied to any
data set with no further information; and realistic, because it should capture
the most likely sources of inaccuracy. These features lead us to making two
assumptions.

The first assumption is that the correlations of experimental uncertainties
across data points are determined much less precisely than the uncertainties
for each data point, which we presume to be exact. This assumption is known
to hold in practice, since the determination of certain correlations --- such
as those for two-point uncertainties defined as the difference between
estimates obtained with two different Monte Carlo generators --- require a
certain amount of guesswork. This fact is occasionally reflected in different
correlation models being presented with the measurement. Therefore we write
\begin{equation}
  A = D A_\text{corr}
  \label{eq:correlation_only}
  \,,
\end{equation}
where $D$ is the $N_{\rm dat} \times N_{\rm dat}$ diagonal matrix of standard
deviations for each data point
\begin{equation}
  D_{ii} = \sqrt{\sum_j^{N_{\rm err}} A_{ij}^2}
  \,.
\end{equation}
We then assume that the covariance matrix provided by or built from the
experiment has the true standard deviations, but correlations (encoded in
$\bar{A}_\text{corr}$ below) may be different from the truth. Analogously to
Eq.~\eqref{eq:abarfromF} we can therefore write
\begin{equation}
  A = D(\bar{A}_\text{corr} + \delta F_\text{corr})
  \,.
\end{equation}
Note that $A_\text{corr}A_\text{corr}^t$ is the covariance matrix of the reduced
differences $(d_i - t_i)/D_{ii}$, hence the analysis carried out in
Sect.~\ref{sec:stability} can be repeated {\it verbatim} for these variables.
Analogously to Eq.~\eqref{eq:deltachi2fromabar}, we can write
\begin{equation}
  \Delta \chi^2 = \norm{A^+_\text{corr}A_\text{corr}} - N_{\rm dat}    
  \,,
\end{equation}
and finally arrive at a stability criterion, similar to
Eq.~\eqref{eq:finalstability}, under the assumption that $D$ is well determined,
\begin{equation}
  \norm{\bar{A}_\text{corr}^+}_2\norm{F_\text{corr}}_F
  \leq
  \frac{1}{\sqrt{2}\delta}
  \,.
  \label{eq:finalstabilitycorr} 
\end{equation}

The second assumption is that $\norm{F_\text{corr}}_F$ is independent of the
number of data points or correlated experimental uncertainties in the
measurement. The model then implies that the prevalent source of inaccuracy in
the correlation matrix concentrates on a subset of data points and originates
from a small number of correlated experimental uncertainties (for example the
correlation of some two-point systematic uncertainties between the most extreme
kinematic bins). While this assumption is a simplification, we find that the
model is effective, as we will discuss in the context of both a toy model (see
Sect.~\ref{subsec:toy_model}) and of a realistic case (see
Sect.~\ref{subsec:fits}). If instead the source of inaccuracy in the correlation
matrix arised from a number of systematic uncertainties that increased, {\it
e.g.}, with the number of data points $N_{\rm dat}$, the regularisation
procedure described below would over-regularise small data sets and
under-regularise large ones when simultaneously applied to a collection of
measurements.

Since $F_\text{corr}$ is a matrix of adimensional coefficients (both units
and magnitude of the data uncertainties are absorbed in $D$), we can simply
set the norm to a constant, {\it e.g.} $\norm{F_\text{corr}}_F = 1/\sqrt{2}$.
Therefore, with the assumptions we have made, the model of uncertainties
required to implement the stability criterion Eq.~\eqref{eq:stabilitysimple}
contains one single adimensional parameter, $\delta$, and the stability
condition is
\begin{equation}
  \norm{\bar{A}_\text{corr}^+}_2 \leq \frac{1}{\delta}
  \label{eq:finalstabilitysimple}
  \,.
\end{equation}
The free parameter $\delta$ characterises the precision of the correlation
matrix. Its optimal value depends on the features of the measurement, and
clearly cannot be obtained from the matrix itself. In the case of PDF
determination, we will obtain it by studying the dependence of global fits
on it, as we will discuss in Sect.~\ref{sec:PDFs}.  

The stability condition Eq.~\eqref{eq:finalstabilitysimple}, together with the
requirements presented at the beginning of this section, lay out a
regularisation procedure. Specifically, Eq.~\eqref{eq:finalstabilitysimple}
implies that the largest singular values of $\bar{A}^+_\text{corr}$ must be bound
by $\delta^{-1}$, and conversely that the smallest singular values of
$\bar{A}_\text{corr}$ must be bound by $\delta$ from below. The requirement that
the regularised matrix gives the same
description as the original one in the directions that do not contribute to
instability implies that the singular vectors with singular values greater than
$\delta^{-1}$ are unchanged. Following Eq.~\eqref{eq:correlation_only}, we write
$\bar{A}$ in terms of the singular value decomposition of $\bar{A}_\text{corr}$,
$\bar{A}_\text{corr} = USV^t$
\begin{equation}
  \bar{A} = DUSV^t
  \label{eq:SVDAbar}
  \,,
\end{equation}
and we can then define the regularised matrix $A_{\rm reg}$ as
\begin{equation}
  \bar{A}_{\rm reg} = DUS_\text{reg}V^t
  \label{eq:regularised_matrix}
  \,,
\end{equation}
where $S_{\rm reg}$ is the matrix of singular values whose non-zero entries are
\begin{equation}
  S_{\text{reg}(ii)}=
  \begin{cases}
    \delta & s_{i}<\delta\\
    s_{i} & \text{otherwise}
  \end{cases}
  \label{eq:regularisation_prescription}
  \,.
\end{equation}

Note that, beside the formulation of the regularisation procedure laid
out above, the stability condition, Eq.~\eqref{eq:finalstabilitysimple},
can be inverted to quickly assess the stability of experimental uncertainties
in a given measurement. We define the condition number $Z$ as
\begin{equation}
  Z = \norm{\bar{A}_\text{corr}^+}_2 = \norm{\bar{A}_\text{corr}}_2^{-1}
  \,.
  \label{eq:Z}
\end{equation}
It follows from Eqs.~\eqref{eq:finalstabilitysimple}
and~\eqref{eq:regularisation_prescription} that, if $Z>\delta^{-1}$, then it is
likely that the precision with which correlations are determined is insufficient
to ensure that they will not alter the expectation value of the $\chi^2$. This
is demonstrated below with a toy model.

We note that the regularisation procedure can apply without modification to
joint matrices constructed from multiple measurements when assuming the same
value of $\delta$ for each of them. For example, if the measurements
are independent, and the joint matrix is block diagonal, with each block
corresponding to the covariance matrix from one measurement, the effect of the
regularisation on the joint matrix is the same as applying it independently to
each of the individual matrices, while the $Z$ condition number will be the
maximum across the measurements. Systematic uncertainties that are shared
between measurements (hence making the joint matrix not completely block
diagonal) also require no change in the procedure.

Finally, we remark that Eq.~\eqref{eq:regularisation_prescription} should work
for any value of $\delta$, even if it has been derived by neglecting terms
of $\mathcal{O}(\delta^2)$ in Eq.~\eqref{eq:deltachi2bound0}. This neglect,
however, may make the interpretation of $\delta^{-1}$ as a measure of the
precision with which correlations need to be known to ensure the stability of
the $\chi^2$ looser for small values of $\delta$.

\subsection{Toy model}
\label{subsec:toy_model}

We now apply the regularisation procedure devised in
Sect.~\ref{subsec:regularisation_procedure} to a toy model which is
representative of a realistic LHC data set. This exercise will show how
inaccuracies in the degree of correlation of uncertainties can undermine
the reliability of the $\chi^2$ as a figure of merit.

The model consists of a data set made of four experimental data points, with a
small uncorrelated statistical uncertainty of size $\epsilon$, equal for each
data point, and one correlated systematic uncertainty of size 1, affecting only
the first three data points. The fourth point also has a systematic uncertainty,
whose correlation with the other points is, however, not precisely known. We
parametrise this lack of knowledge in terms of the variable $x$, and write the
systematic uncertainty on the fourth point as a fluctuation, by an amount $x$,
with respect to the other systematic uncertainty of size 1. We assume that the
total variance is known. Note that this is consistent with the assumptions made
in Sect.~\ref{subsec:regularisation_procedure}: correlations can fluctuate (in a
way that, in the model, is parametrised by $x$), while variances remain fixed.

The matrix of uncertainties describing this toy model is
\begin{equation}
  A(x) = \left(
  \begin{matrix}
    \epsilon & 0 & 0 & 0 & 1 & 0\\
    0 & \epsilon & 0 & 0 & 1 & 0\\
    0 & 0 & \epsilon & 0 & 1 & 0\\
    0 & 0 & 0 & \epsilon & 1 - x & \sqrt{1 - \left(1 - x\right)^{2}}\\
  \end{matrix}
  \right)
  \label{eq:A_toy}
  \,.
\end{equation}
By fixing the variance due to the systematics to 1 we let the parameter
$\epsilon\ll 1$  control the relative size of the uncorrelated to correlated
uncertainties. The parameter $x$ can take values in the interval $[0, 2]$:
$x=0$ corresponds to the case in which the systematic uncertainty on the fourth
data point is fully correlated with that of the other data points; $x=1$
corresponds to the case of full decorrelation; and $x=2$ corresponds to the
case of full anti-correlation. 

We now consider the situation in which the correlation is (inaccurately)
estimated to be maximal, that is $x=\bar{x}=0$. This inaccuracy is encoded in
the matrix of uncertainties $\bar{A} = A(\bar{x})$:
\begin{equation}
  \bar{A} = \left(
  \begin{matrix}
    \epsilon & 0 & 0 & 0 & 1 & 0\\
    0 & \epsilon & 0 & 0 & 1 & 0\\
    0 & 0 & \epsilon & 0 & 1 & 0\\
    0 & 0 & 0 & \epsilon & 1 & 0\\
  \end{matrix}
  \right)
  \label{eq:Abar_toy}
  \,.
\end{equation}
According to Eq.~\eqref{eq:badchi2asfrob}, the expectation value of the
$\chi^2$ given $\bar A$ is
\begin{equation}
  \expval{\bar{\chi}^2}(x)
  =
  \norm{\bar{A}^+ A(x)}_F^2
  =
  4 + \frac{6x}{\epsilon^2(\epsilon^2+4)}
  \,,
  \label{eq:toy_bad_chi2}
\end{equation}
which has to be compared with the true expectation value given $A(x)$, see
Eq.~\eqref{eq:chi2asfrob}:
\begin{equation}
  \expval{\chi^2_{\rm true}}(x)
  = \norm{A^+ A(x)}_F^2
  = N_{\rm dat} = 4
  \,.
  \label{eq:toy_true_chi2}
\end{equation}

\begin{figure}[!t]
    \centering
    \includegraphics[width=0.49\columnwidth]{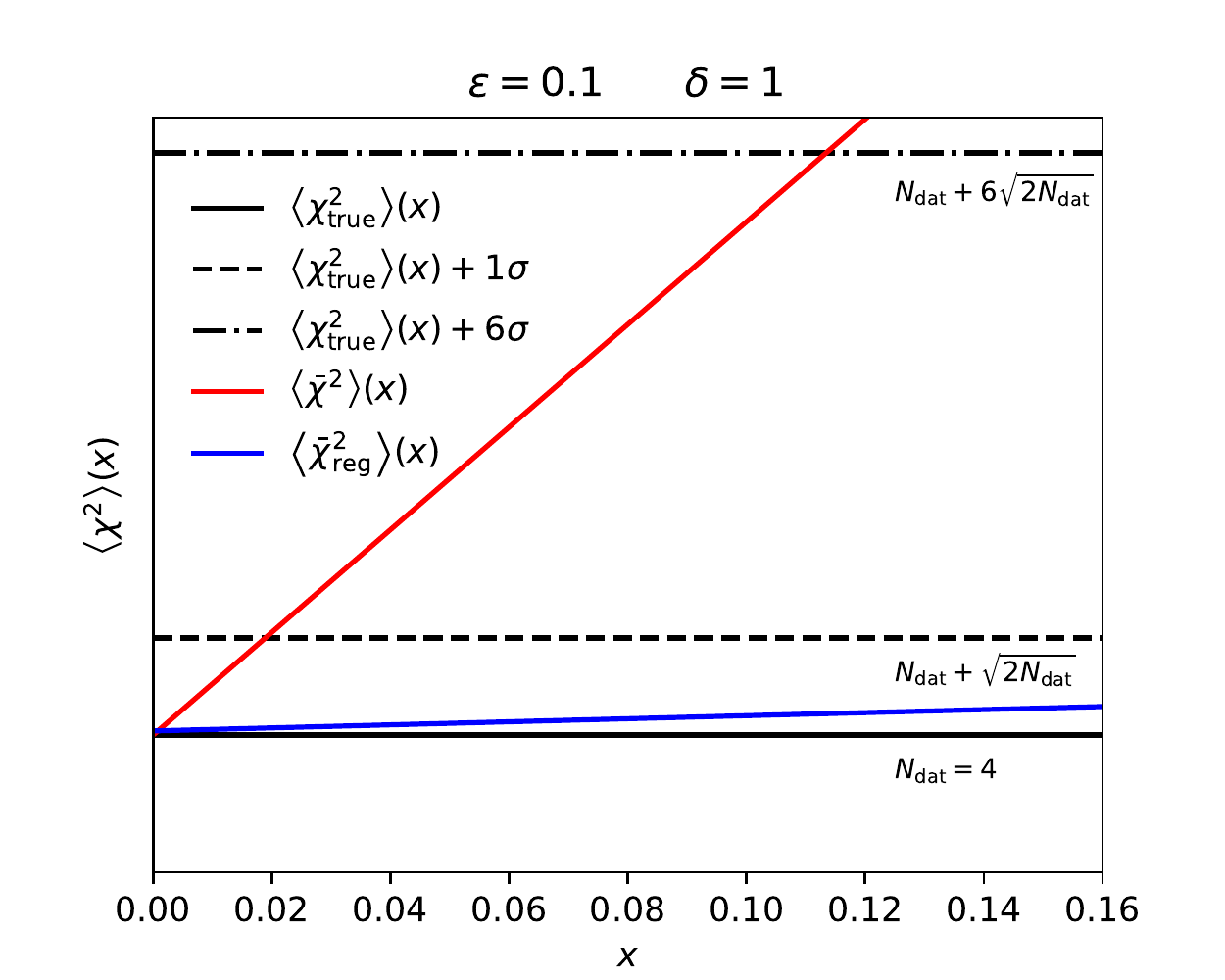}
    \includegraphics[width=0.49\columnwidth]{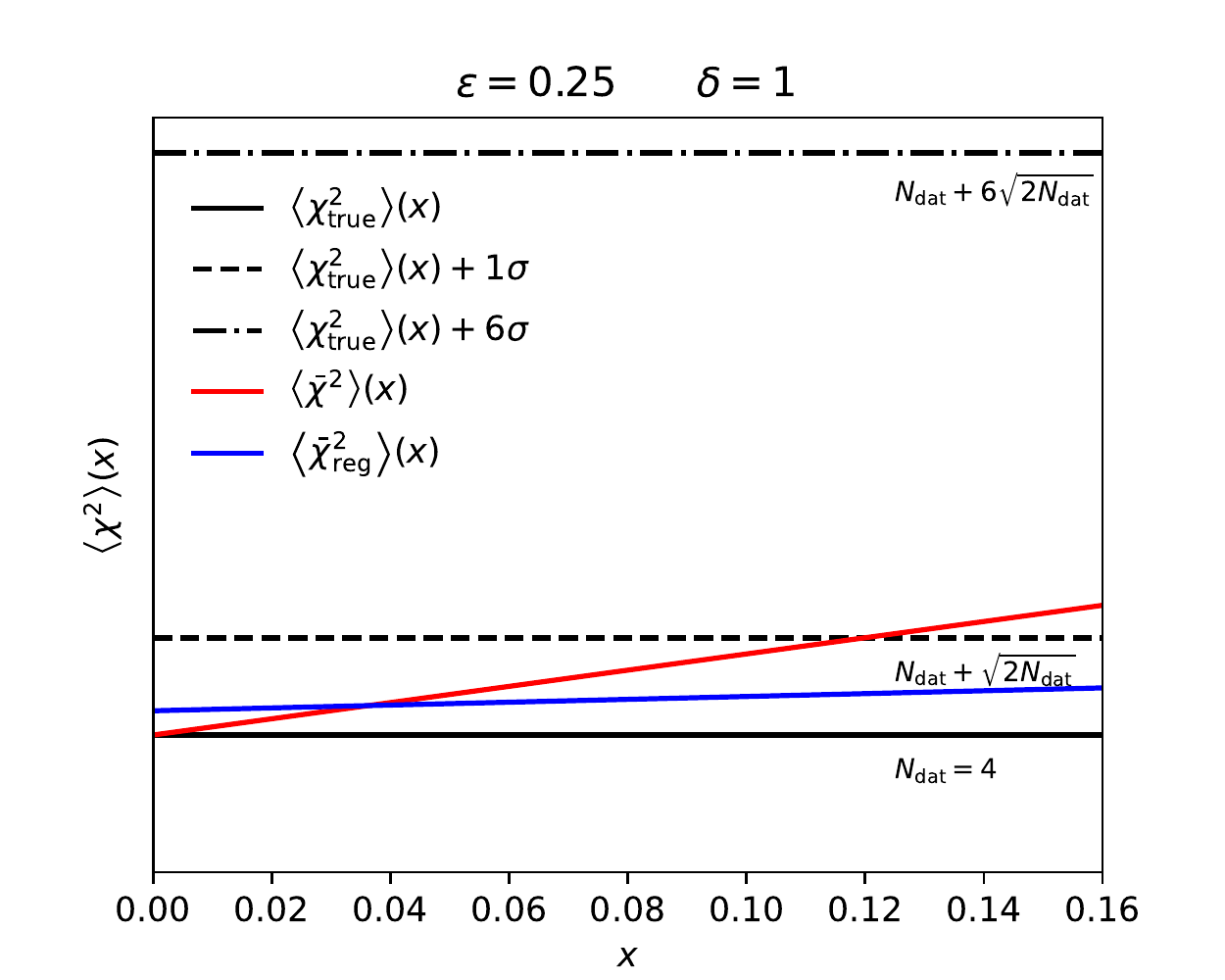}\\
    \vspace{0.5cm}
    \caption{The expectation value of the $\chi^2$, as a function of the
      variable $x$, in the toy model, for two values of the parameter
      $\epsilon$: $\epsilon=0.1$ (left) and $\epsilon=0.25$ (right). We show:
      the true expectation value of Eq.~\eqref{eq:toy_true_chi2},
      $\expval{\chi^2_{\rm true}}$ (with one and six standard deviations for
      reference), given the matrix of uncertainties $A$, Eq.~\eqref{eq:A_toy};
      the expectation value of Eq.~\eqref{eq:toy_bad_chi2},
      $\expval{\bar\chi^2}$, given the inaccurate matrix of uncertainties
      $\bar{A}$, Eq.~\eqref{eq:Abar_toy}; and the expectation value of
      Eq.~\eqref{eq:toy_reg_chi2}, $\expval{\bar\chi^2_{\rm reg}}$, given the
      matrix of uncertainties $\bar{A}_{\rm reg}$, Eq.~\eqref{eq:Areg_toy},
      obtained after applying the regularisation procedure with $\delta=1$.}
    \label{fig:toychi2}
\end{figure}

The situation is depicted in Fig.~\ref{fig:toychi2}, where the curves obtained
with either Eq.~\eqref{eq:toy_bad_chi2} or Eq.~\eqref{eq:toy_true_chi2} are
contrasted as a function of the true (unknown) variable $x$. We consider two
illustrative values of the model parameter $\epsilon$, $0.1$ and $0.25$,
that correspond to the situation in which the uncorrelated statistical
uncertainty is equal, respectively, to 10\% or 25\% of the correlated
systematic uncertainty. These values reflect the relative ratio of uncorrelated
to correlated uncertainties in realistic current and future LHC measurements.
As is apparent from Fig.~\ref{fig:toychi2}, the incorrect estimation of
$\bar{x}$ leads to a large deviation of the expectation value of the $\chi^2$
from its true value. The smaller the value of $\epsilon$, the larger the
deviation. For example, for a value of $\epsilon$ equal to $0.25$, it is
sufficient that the true value of $x$ is $0.12$ instead of zero to run afoul of
the stability criterion of Eq.~\eqref{eq:stabilitysimple}. For $\epsilon=0.1$,
the true value of $x$ can be as small as $0.02$ to encounter a similar
instability.

We now apply the regularisation procedure devised in
Sect.~\ref{subsec:regularisation_procedure}. We first write the matrix
$\bar{A}$, Eq.~\eqref{eq:Abar_toy}, in terms of the matrices $D$ and
$\bar{A}_{\rm corr}$, as per Eq.~\eqref{eq:correlation_only}, which read
\begin{equation}
  D = \sqrt{1+\epsilon^2}\,I_{4\times 4}
  \qquad
  \text{and}
  \qquad
  \bar{A}_{\rm corr} = \frac{1}{\sqrt{1+\epsilon^2}}\,\bar{A}
  \label{eq:D_and_Abarcorr_toy}
  \,.
\end{equation}
The matrix of singular values for $\bar{A}_{\rm corr}$ is
\begin{equation}
  S = \frac{1}{\sqrt{1+\epsilon^2}}\left(
  \begin{matrix}
    \epsilon & 0 & 0 & 0 & 0 & 0\\
    0 & \epsilon & 0 & 0 & 0 & 0\\
    0 & 0 & \epsilon & 0 & 0 & 0\\
    0 & 0 & 0 & \sqrt{4+\epsilon^2} & 0 & 0\\
  \end{matrix}
  \right)
  \label{eq:singular_values_toy}
  \,.
\end{equation}
We denote the first three singular values as
$s_{1,2,3}=\epsilon/\sqrt{1+\epsilon^2}$ and the fourth one as
$s_{4}=\sqrt{4+\epsilon^2}/\sqrt{1+\epsilon^2}$, and note that
$0<s_{1,2,3}<s_4<2$ for any value of $\epsilon>0$. We then apply the
regularisation prescription given by
Eqs.~\eqref{eq:SVDAbar}-\eqref{eq:regularisation_prescription},
by choosing $s_{1,2,3}<\delta^{-1}<s_4$. The regularised matrix of singular values
therefore reads
\begin{equation}
  S_{\rm reg} = \left(
  \begin{matrix}
    \delta^{-1} & 0 & 0 & 0 & 0 & 0\\
    0 & \delta^{-1} & 0 & 0 & 0 & 0\\
    0 & 0 & \delta^{-1} & 0 & 0 & 0\\
    0 & 0 & 0 & \frac{\sqrt{4+\epsilon^2}}{\sqrt{1+\epsilon^2}} & 0 & 0\\
  \end{matrix}
  \right)
  \label{eq:singular_values_reg_toy}
  \,,
\end{equation}
and the regularised matrix of uncertainties
\begin{equation}
  \bar{A}_{\rm reg} = \left(
  \begin{matrix}
    a & b & b & b & 1 & 0\\
    b & a & b & b & 1 & 0\\
    b & b & a & b & 1 & 0\\
    b & b & b & a & 1 & 0\\
  \end{matrix}
  \right)
  \,,
  \label{eq:Areg_toy}
\end{equation}
where
\begin{equation}
  a = \frac{1}{4}\left(\epsilon + 3\delta^{-1}\,\sqrt{1+\epsilon^2}\right)
  \qquad
  \text{and}
  \qquad
  b = \frac{1}{4}\left(\epsilon - \delta^{-1}\,\sqrt{1+\epsilon^2}\right)
  \,.
  \label{eq:ab_toy}
\end{equation}
The expected value of the $\chi^2$ is finally
\begin{equation}
\expval{\bar{\chi}^2_{\rm reg}}(x)
  =
  \norm{\bar{A}^+_{\rm reg} A(x)}_F^2
  =
  4
  +
  6x\left( \frac{\delta^2}{1 + \epsilon^2} - \frac{1}{4+\epsilon^2}\right)
  +
  \frac{12\,\delta^2\epsilon^2}{1 + \epsilon^2}
  \,.
  \label{eq:toy_reg_chi2}  
\end{equation}

The expression in Eq.~\eqref{eq:toy_reg_chi2} is compared to those in
Eqs.~\eqref{eq:toy_bad_chi2}-\eqref{eq:toy_true_chi2} in Fig.~\ref{fig:toychi2}
for the value $\delta=1$. We note that this value fulfils the requirement
$s_{1,2,3}<\delta^{-1}<s_4$ for any value of the parameter $\epsilon$. As is
apparent from Fig.~\ref{fig:toychi2}, the regularisation procedure successfully
achieves the goal for which it was devised: the expectation value of the
regularised $\chi^2$, $\expval{\chi^2_{\rm reg}}$, does not differ from the true
expectation value, $\expval{\chi^2}_{\rm true}$, by more than one standard
deviation of the $\chi^2$ distribution for any value of $x$.

The optimal value of $\delta$ should be determined on a case-by-case basis
depending on the precision with which $x$ is known. This is the topic that we
will investigate in the next section in the context of PDF determination.

We now turn our attention to the situation where we can make further assumptions
on the uncertainties in the determination of the correlation structure, for
example when having access to additional information during the experimental
analysis. In that case it might be advisable to study the effects on stability
of various modelling choices, and the corresponding regularisation, in a more
refined way than the one described in
Sec.~\ref{subsec:regularisation_procedure}, where we strived for generality. We
simulate this situation by assuming a specific prior for the value of the $x$
parameter. We choose that prior to be a beta distribution with support in $x \in
[0, 1]$ and such that $x=0$ is the mode value. Specifically,
\begin{equation}
  \label{eq:xprior}
  x \sim \operatorname{Beta}(1, 5)
  \, ,
\end{equation}
which corresponds to the probability density
\begin{equation}
  f_x(\xi) = 5(1 - \xi)^4
  \, .
\end{equation}
Our discussion implies that even though $x=0$ is the most likely value,
analyses using it are subject to instabilities. We can quantify this by
computing the expected error in the $\chi^2$ we would incur when assuming a
particular value of $x$ and when averaging over the distribution of possible
values:
\begin{equation}
  \label{eq:integx}
  \expval{\Delta \chi^2}(x) = \int_0^1   \abs{\norm{\bar{A}^+(\xi) A(x)}_F^2 - N}\, f_x(\xi)\text{d}\xi
  \, .
\end{equation}

\begin{figure}[!t]
    \centering
    \includegraphics[width=0.70\columnwidth]{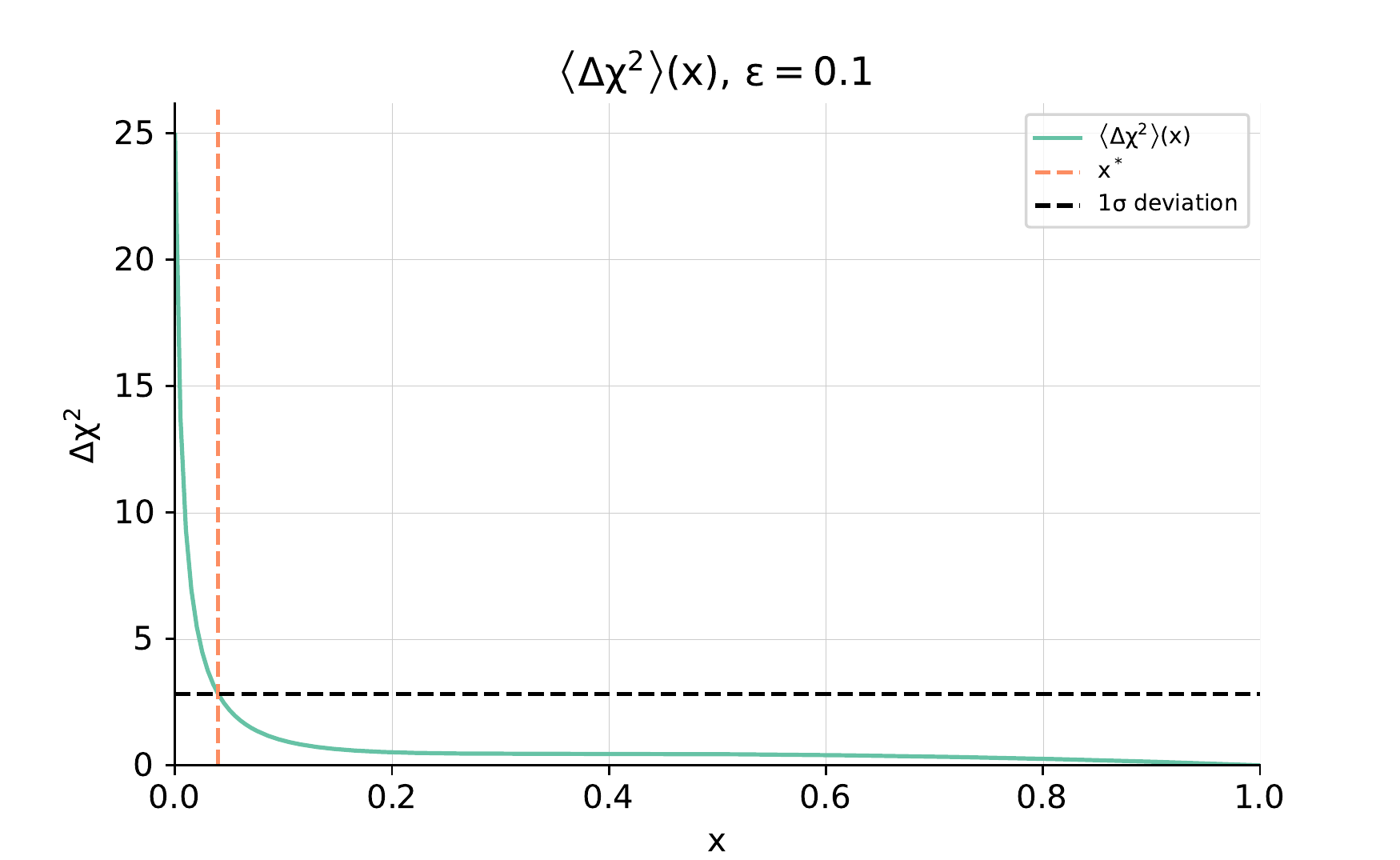}
    \vspace{0.5cm}
    \caption{Stability of the toy model with additional assumptions on the value
    of the correlation. The green curve shows the deviation in $\chi^2$ averaged
    over the assumed prior of the $x$ parameter, Eq.~\eqref{eq:integx}. The
    black dashed horizontal line marks the limit from the stability criterion
    Eq.~\eqref{eq:stabilitysimple}. The orange dashed vertical line, in the
    intersection, shows the most likely value of $x$ that fulfills the stability
    criterion, Eq.~\eqref{eq:bestx}.
    }
    \label{fig:toyexp}
\end{figure}

We represent $\expval{\Delta \chi^2}_x(x)$ in Fig.~\ref{fig:toyexp}, where we
have set $\epsilon=0.1$. The comparison with the limit imposed by the stability
criterion Eq.~\eqref{eq:stabilitysimple}, also displayed in
Fig.~\ref{fig:toyexp}, shows that presenting the covariance matrix with values
too close to the most likely value of the correlation under the prior
yields large instabilities that would hamper the subsequent analysis.
Selecting the most likely value that satisfies the stability criterion
\begin{equation}
  \label{eq:bestx}
  x^*
  =
  \operatorname*{argmax}_{\xi: \left\langle\Delta \chi^2\right\rangle(\xi) \leq \sqrt{2N}} f_x(\xi)
\end{equation}
may be a way to decide the value of the correlation with which to present the
covariance matrix.
This would correspond to $x \approx 0.04$ under the settings presented here.
Note that this small correction is consistent with the assumed knowledge of $x$,
Eq.~\eqref{eq:xprior}, but it would notably increase the accuracy of $\chi^2$
computations using the covariance matrix.

The obvious disadvantage of this analysis is the difficulty of obtaining
estimates for the covariance matrix parameters such as Eq.~\eqref{eq:xprior}.
These are unattainable outside the experimental collaborations responsible for
the analysis and presumably challenging within. However, it may be useful to
assess  and refine correlation models internally.  The regularisation procedure
presented in Sec.~\ref{sec:regularisation} and applied to the toy model in
Eq.~\eqref{eq:toy_reg_chi2} is indicated for the more common situation where
such detailed information is missing. We demonstrate its usage for the problem
of PDF determination next.

%% file: sec-PDFs.tex
\section{Determining PDFs with a regularised data set}
\label{sec:PDFs}

In this section, we apply the regularisation procedure devised in
Sect.~\ref{sec:regularisation} to a data set utilised for PDF determination.
This is a particular problem relevant to LHC precision physics that relies on
the $\chi^2$ as a figure of merit. We first discuss how the regularisation
procedure can be applied to characterise the data set that enters a given PDF
determination. We then show how PDFs change if the nominal data set is replaced
by a suitably regularised one, and study their dependence on the regularisation
parameter $\delta$. We finally investigate how the regularisation procedure
performs in comparison to the correlation models provided with the measurements
in the few cases in which these are available. All of our investigations are
performed in the framework of the recent NNPDF4.0 PDF
determination~\cite{NNPDF:2021njg}.

\subsection{Characterising the data set}
\label{subsec:characterisation}

The NNPDF4.0 data set is the widest data set used for PDF determination to
date. It consists of legacy fixed-target and collider deep-inelastic scattering
and fixed-target Drell--Yan measurements, and of a wide range of measurements
for various production processes in proton--proton collisions at the LHC. These
include both Run I and Run II measurements and make about 30\% of the NNPDF4.0
data set. Experimental uncertainties are typically of the order of few percent,
the largest part of which is made of correlated systematic uncertainties. A
detailed description of the NNPDF4.0 data set is provided in Sect.~2
of~\cite{NNPDF:2021njg}.

Here we take a closer look at the LHC measurements that are part of the
NNPDF4.0 data set, and in particular scrutinise the matrix of uncertainties
of each measurement that contains more than one data point. The goal is to
identify the measurements for which an inaccurate estimation of experimental
correlations may significantly affect their $\chi^2$. To this purpose, for each
measurement, we compute the condition number $Z$, Eq.~\eqref{eq:Z}, apply the
regularisation procedure delineated in Sect.~\ref{sec:regularisation}
for different values of the parameter $\delta$, and evaluate how much the
regularised covariance matrix differs from the nominal one. This piece of
information is collected in Table~\ref{tab:differences}, where we indicate,
for each LHC measurement included in the NNPDF4.0 data set, its reference and
the condition number $Z$; we also indicate the maximum relative difference of
the variances $\Delta\sigma_r$ and the maximum absolute difference of the
correlation $|\Delta\rho|$ computed between the nominal data set and the data
set regularised with $\delta^{-1}=1,2,3,4,5,7$. Blank spaces indicate that
$\Delta\sigma_r=|\Delta\rho|$=0, that is the regularisation procedure does not
alter the nominal covariance matrix. We make two remarks.

\begin{table}[!t]
  \scriptsize
  \centering
  \renewcommand{\arraystretch}{1.4}
  \input{tables/tab-differences.tex}\\
  \vspace{0.5cm}
  \caption{The LHC measurements included in the NNPDF4.0 data
    set~\cite{NNPDF:2021njg}. For each measurement we indicate its reference,
    the condition number $Z$ of the corresponding experimental covariance
    matrix, Eq~\eqref{eq:Z}, and the maximum relative difference of the
    variances $\Delta\sigma_r$ (in percent) and the maximum absolute difference
    of the correlation $|\Delta\rho|$ computed between the nominal data set and
    the data set obtained by applying the regularisation procedure delineated
    in Sect.~\ref{sec:regularisation} for $\delta^{-1}=1,2,3,4,5,7$. Blank
    spaces indicate that $\Delta\sigma_r=|\Delta\rho|=0$, that is the
    regularisation procedure does not alter the nominal covariance matrix.
    For ATLAS $W,Z$~7~TeV, CC and CF stand, respectively, for central-central
    and central-forward rapidity selections. We omit the data sets with a
    single data point.}
  \label{tab:differences}
\end{table}

First, one can single out the data sets for which an inaccurate estimation of
experimental correlations may be of concern in a PDF fit. These are the data
sets with the largest value of the condition number $Z$. If these data sets
turn out to also have an unsatisfactory $\chi^2$ in the fit, then additional
investigations are needed to establish whether this is due solely to inaccurate
experimental correlations, solely to inaccurate theoretical predictions, or
to a combination of both. Conversely, if a data set has a low condition number
$Z$ but a large value of the $\chi^2$, the large value of the $\chi^2$ is
reasonably due to genuine inconsistencies between the data set and theory
predictions. These considerations may help determine the optimal data set
utilised as input to PDF determination, as done for the NNPDF4.0 parton set
(see in particular Sect.~4.2 in~\cite{NNPDF:2021njg}).

Second, one can determine the optimal value of the regularisation parameter
$\delta$ in such a way that variances and correlations are not modified too
much by the regularisation procedure in comparison to the nominal values.
In this respect, inspection of Table~\ref{tab:differences} reveals that
regularising the NNPDF4.0 data set with $\delta^{-1}=1$ or $\delta^{-1}=2$ is too
aggressive, in that it leads to an increase of variances by an amount between
10\% and 90\%, and a variation of correlations between 0.1 and 0.5, depending
on the data set. These figures are reduced, respectively, below 10\% and 0.1
for $\delta^{-1}=3$ and even further, to a few percent and below 0.05 for
$\delta^{-1}=4$
and $\delta^{-1}=5$. Higher values of $\delta$ alter the nominal data set only
minimally. As expected, the data sets associated to the highest condition
number $Z$ are those that are generally most affected by the regularisation
procedure, in that they display the largest variation of variances and
correlation; they also remain sensitive to the regularisation procedure even if
a modest amount of regularisation (that is, a high value of $\delta^{-1}$) is
applied. In the next section we shall see how these variations affect a fit of
PDFs. 

\begin{figure}[!t]
  \centering
  \includegraphics[width=0.48\textwidth,clip=true,trim=3cm 0 0.5cm 0]{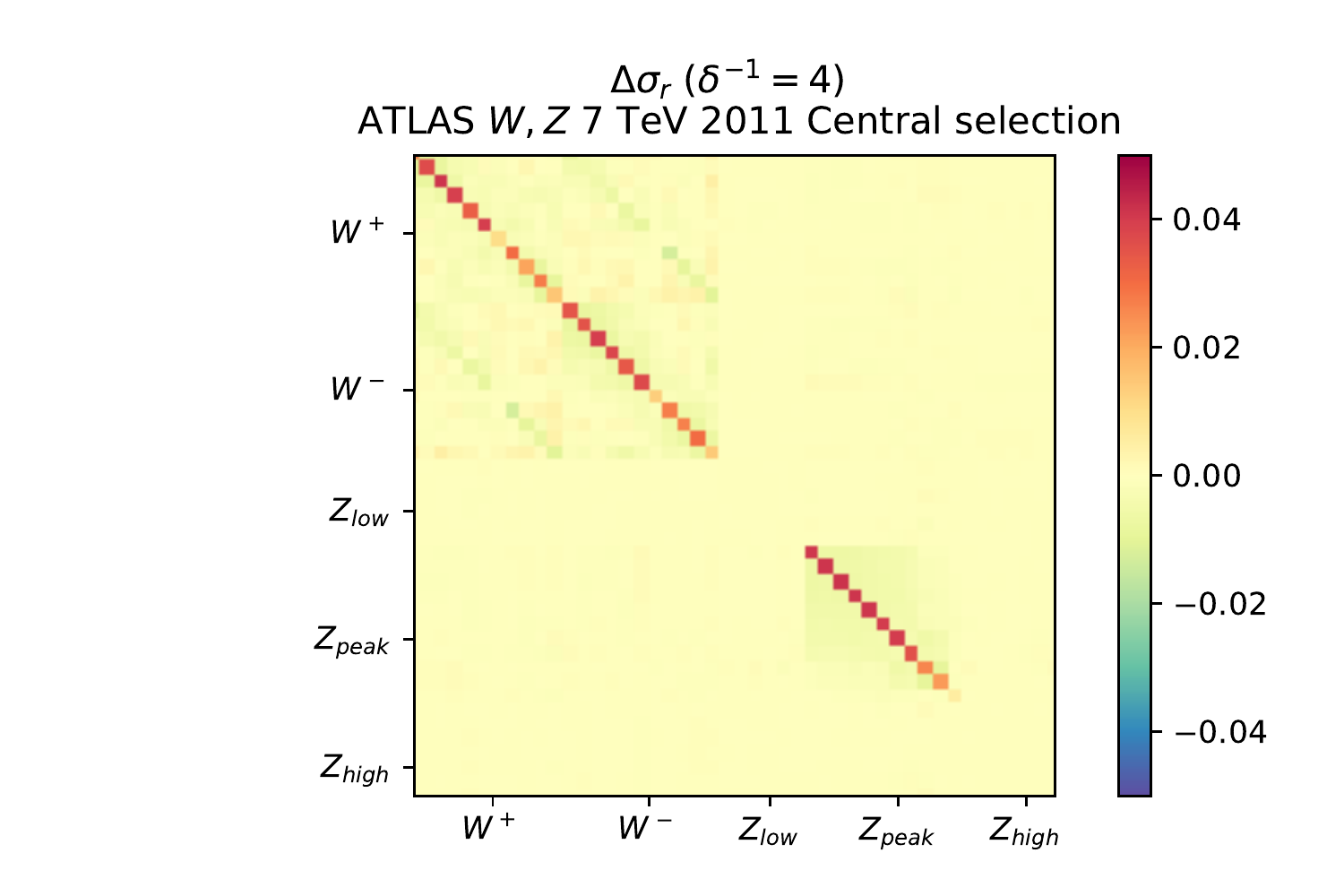}
  \includegraphics[width=0.48\textwidth,clip=true,trim=3cm 0 0.5cm 0]{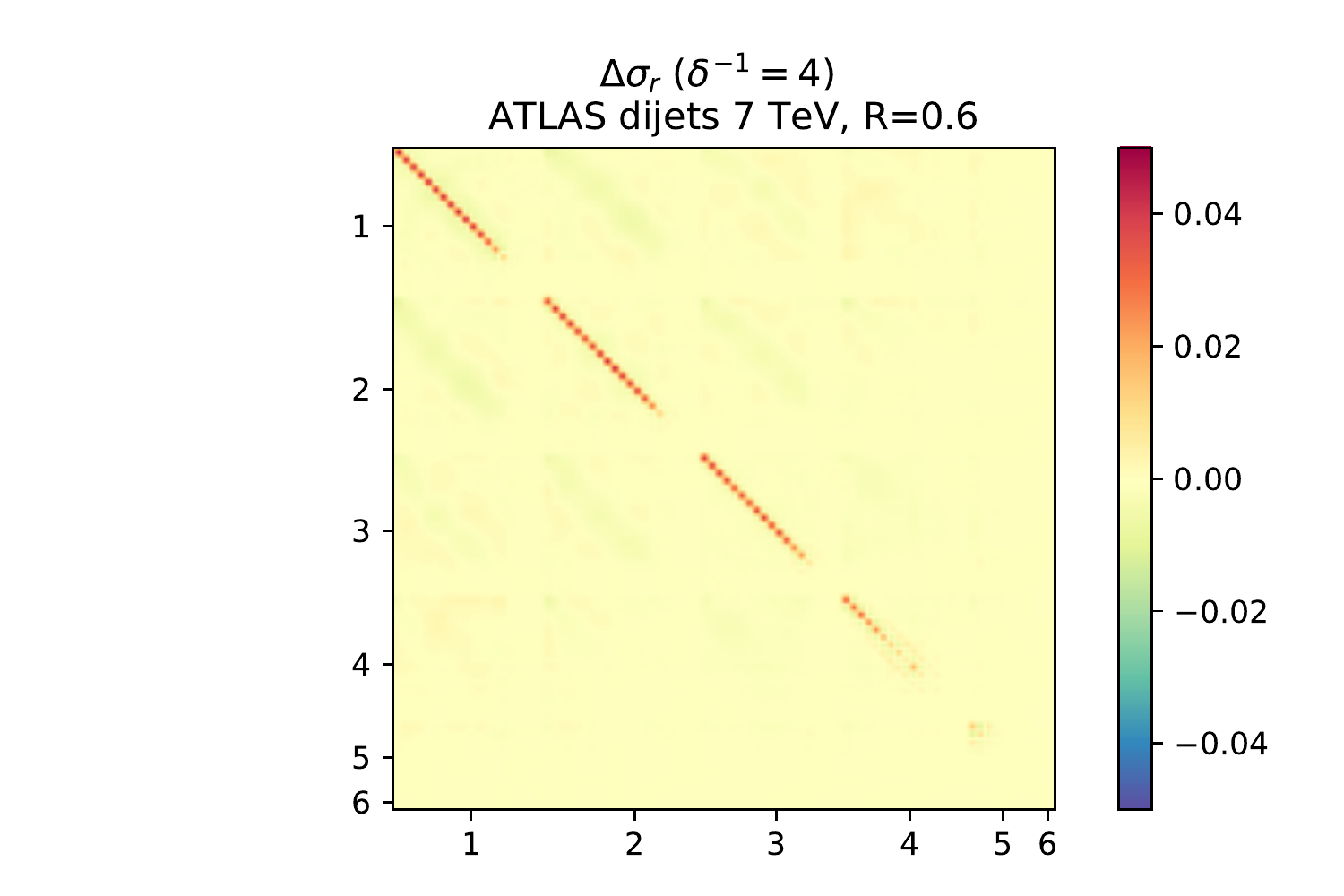}\\
  \includegraphics[width=0.48\textwidth,clip=true,trim=3cm 0 0.5cm 0]{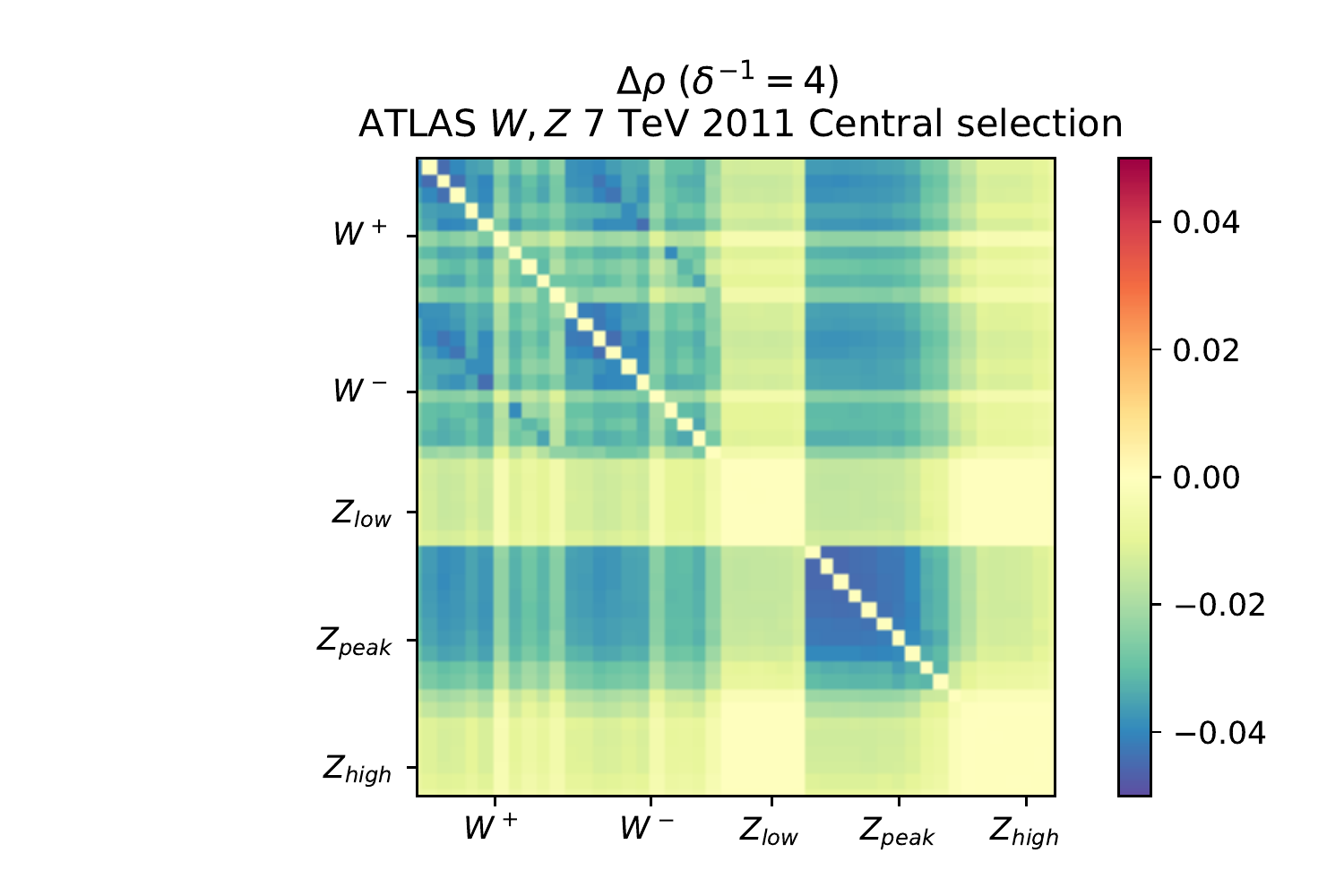}
  \includegraphics[width=0.48\textwidth,clip=true,trim=3cm 0 0.5cm 0]{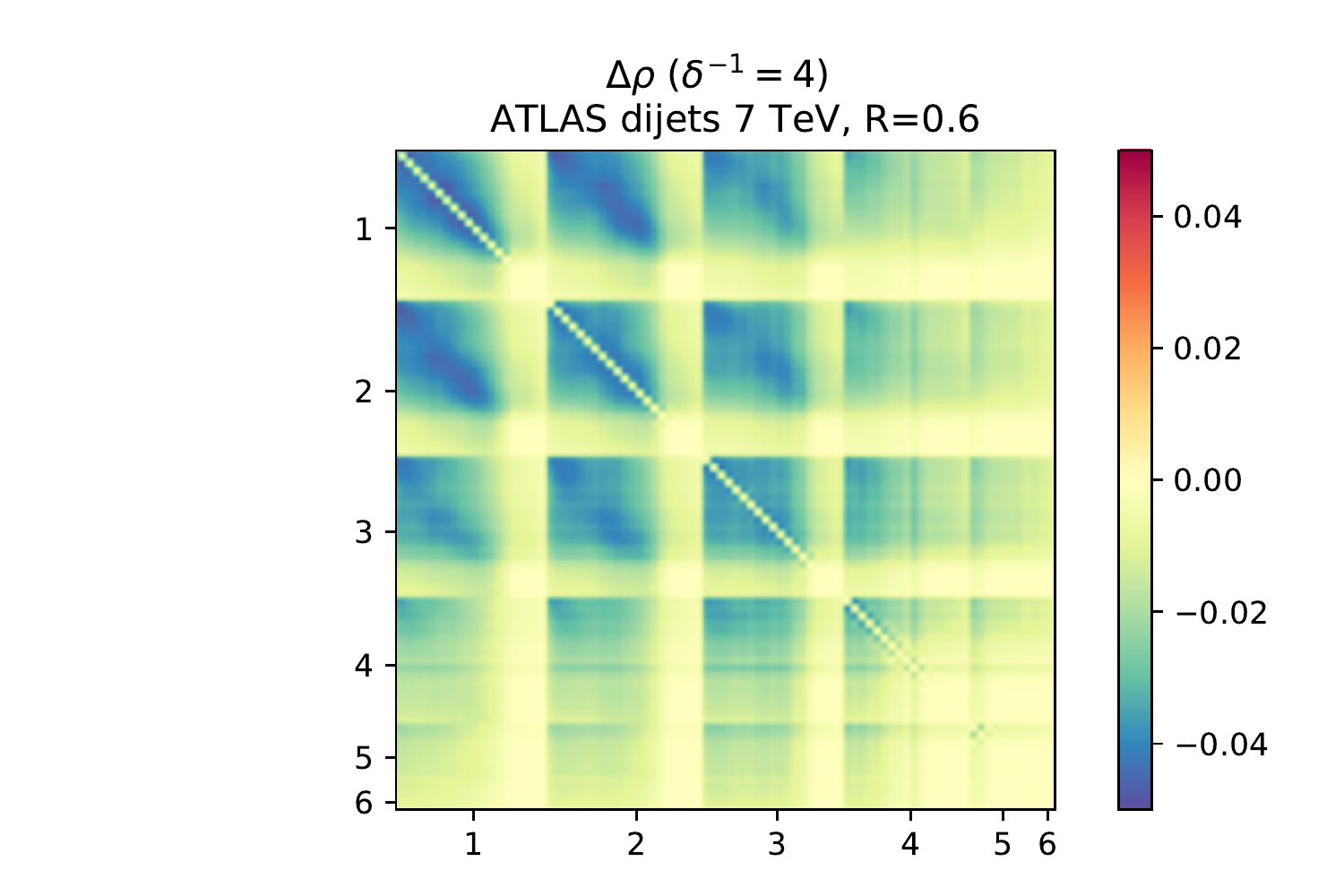}\\
  \vspace{0.5cm}
  \caption{The relative difference of the covariance matrix $\Delta\sigma_r$
    (top) and the difference of the correlation matrix
    $\Delta\rho$ (bottom) for each of their elements, computed between the
    regularised data sets and the data set nominal with $\delta^{-1}=4$. We show
    results for the two measurements in the NNPDF4.0 data set that have the
    largest value of $Z$, see Table~\ref{tab:differences}: ATLAS
    $W,Z$~7~TeV~CC~\cite{ATLAS:2016nqi} (left) and ATLAS
    dijets~R=0.6~7~TeV~\cite{ATLAS:2013jmu} (right). For
    ATLAS $W,Z$~7~TeV~CC, we indicate the bins, differential in the rapidity of
    the lepton, $\eta$, corresponding to $W^+$, $W^-$ and $Z$ production (the
    latter in three kinematic regions); for ATLAS dijets~R=0.6~7~TeV we
    indicate the bins, differential in the invariant mass of the dijet,
    $m_{12}$, corresponding to the six measured intervals of the absolute
    rapidity difference of the two leading jets, $|y^*|$, see text for details.}
  \label{fig:covmatrices}
\end{figure}

Among all of the LHC data sets collected in Table~\ref{tab:differences}, we
single out the two measurements that are associated to large values of $Z$ and
$\chi^2$ (see Table~\ref{tab:chi2}) at the same time:
ATLAS $W,Z$~7~TeV~CC~\cite{ATLAS:2016nqi} and ATLAS
dijets~R=0.6~7~TeV~\cite{ATLAS:2013jmu}. They are representative of extreme
cases in which small inaccuracies in the determination of experimental
correlations can have a large impact on the computation of the $\chi^2$. Indeed
these data sets have been the subject of much scrutiny~\cite{ATLAS:2016nqi,
  Hou:2019efy,Bailey:2020ooq,NNPDF:2021njg}.
With a value of $Z$ of order 10, it means that correlations must be estimated
with an absolute uncertainty of roughly less than 0.1 in order to ensure that
they make the $\chi^2$ fluctuate by less than one standard deviation. If the
correlation between two bins is estimated to be $1.0$ while its real value is
instead $0.9$, one can expect the $\chi^2$ to deviate significantly (by more
than one standard deviation) from unity, even if there is good consistency
between experimental central values and theoretical expectations.

Note that other data sets may have a large value of $Z$, {\it e.g.}
CMS $Z$ $p_T$~8~TeV~\cite{CMS:2015hyl}, but not an anomalously large $\chi^2$
(see Table~\ref{tab:chi2}). While our decorrelation procedure will also
affect these data sets, as seen in Table~\ref{tab:differences}, we do not
consider them in the following discussion.

In Figure~\ref{fig:covmatrices} we show how the regularisation procedure
described in Sect.~\ref{subsec:regularisation_procedure} affects the covariance
and correlation matrices of the two data sets singled out above. Specifically,
we show the relative difference of the covariance matrix $\Delta\sigma_r$ and
the difference of the correlation matrix $\Delta\rho$ for each of their
elements, computed between the nominal data sets and the data set regularised
with $\delta^{-1}=4$. For ATLAS $W,Z$~7~TeV~CC, we indicate the bins,
differential in the rapidity of the lepton, $\eta$, corresponding to $W^+$,
$W^-$ and $Z$ production (the latter in three kinematic regions); for ATLAS
dijets~R=0.6~7~TeV, we indicate the bins, differential in the invariant mass of
the dijet, $m_{12}$, corresponding to the six measured intervals of the
absolute rapidity difference of the two leading jets, $|y^*|$:
$0.0\leq |y^*|\leq 0.5$; $0.5\leq |y^*|\leq 1.0$;
$1.0\leq |y^*|\leq 1.5$; $1.5\leq |y^*|\leq 2.0$; $2.0\leq |y^*|\leq 2.5$;
and $2.5\leq |y^*|\leq 3.0$. As already noted, differences are small and do not
exceed $5\%$ for variances and 0.05 for correlations. These variations seem
very reasonable to us; their effect, as well as that induced by larger (smaller)
variations corresponding to more (less) aggressive regularisation will be
investigated next.

\subsection{Fitting PDFs}
\label{subsec:fits}

We now study the sensitivity of PDF determination to the regularisation
procedure. To this purpose, we perform a series of fits, all based on the
experimental, theoretical, and methodological input that enters the default
next-to-next-to-leading order (NNLO) NNPDF4.0 parton set
(see~\cite{NNPDF:2021njg} for details), in which we regularise the data set.
Specifically, we perform six fits in each of which we consider a different
amount of regularisation, namely $\delta^{-1}=1,2,3,4,5,7$. All the fits are
made of $N_{\rm rep}=100$ Monte Carlo replicas. Note that these fits are different
from those presented in Sect.~8.7 of~\cite{NNPDF:2021njg}: here the
regularisation procedure is applied to the NNPDF4.0 data set as a whole (and
indeed to the total covariance matrix), while there it was applied only to a
specific measurement (that was part of the NNPDF4.0 data set or not) at a time.

In Table~\ref{tab:chi2} we display the value of the $\chi^2$ per data point,
$\chi^2/N_{\rm dat}$, for each of these fits, and compare it to that of the
NNLO NNPDF4.0 default fit. Deep-inelastic scattering, fixed-target
Drell--Yan, and Tevatron Drell-Yan measurements, which are mostly unaffected by
the regularisation procedure, are all aggregated; ATLAS, CMS and LHCb
measurements are instead displayed individually. The total values (for each
experiment and for the total data set) are also shown, as well as the
corresponding number of data points.

\begin{table}[!p]
  \scriptsize
  \centering
  \renewcommand{\arraystretch}{1.4}
  \input{tables/tab-chi2.tex}\\
  \vspace{0.5cm}
  \caption{The number of data points, $N_{\rm dat}$, and the $\chi^2$ per data
    point, $\chi^2/N_{\rm dat}$, for the NNPDF4.0 NNLO baseline fit and for
    each of the fits performed with the regularisation procedure delineated
    in Sect.~\ref{sec:regularisation} for $\delta^{-1}=1,2,3,4,5,7$.}
  \label{tab:chi2}
\end{table}

\begin{figure}[!p]
  \centering
  \includegraphics[width=0.48\textwidth]{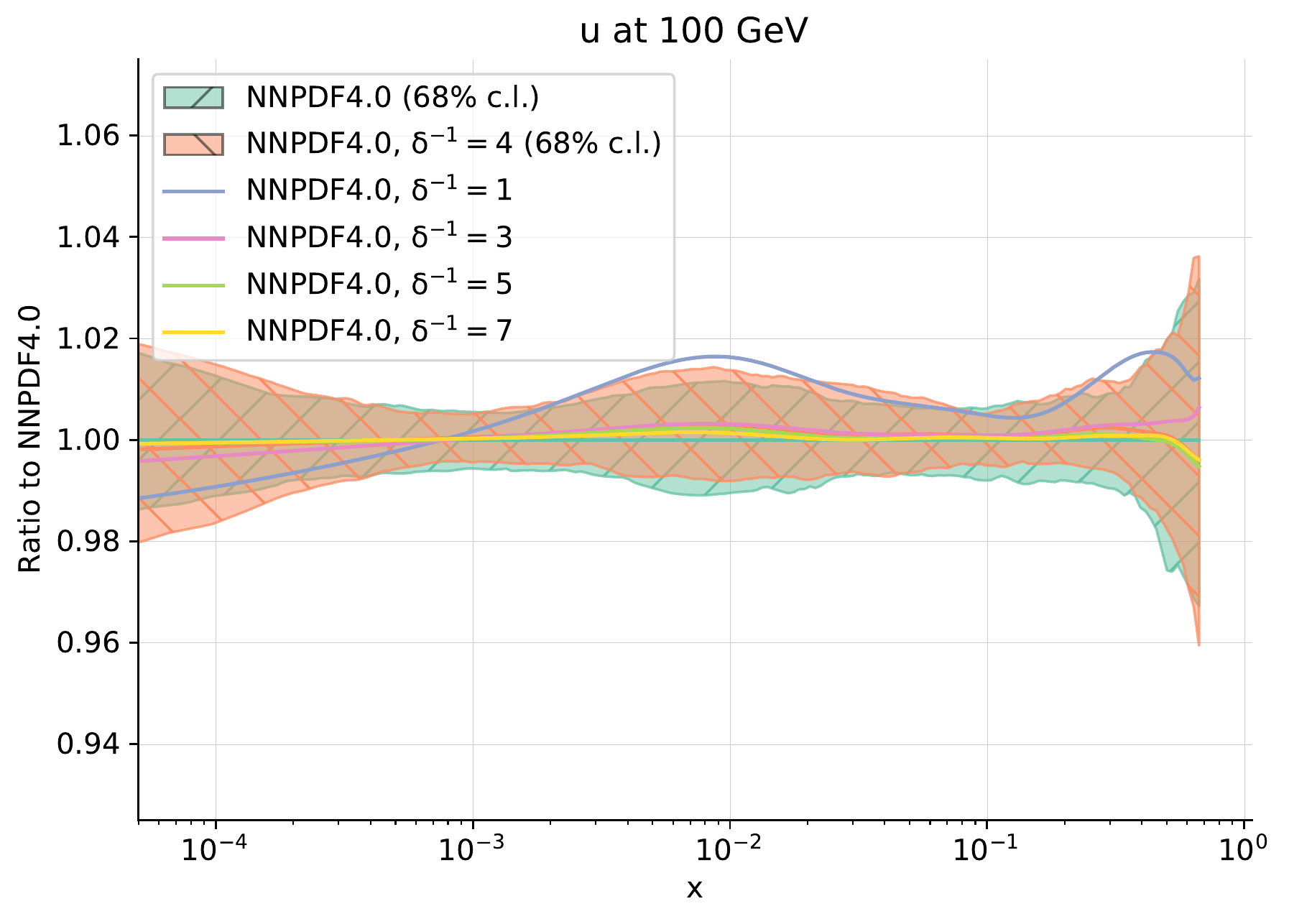}
  \includegraphics[width=0.48\textwidth]{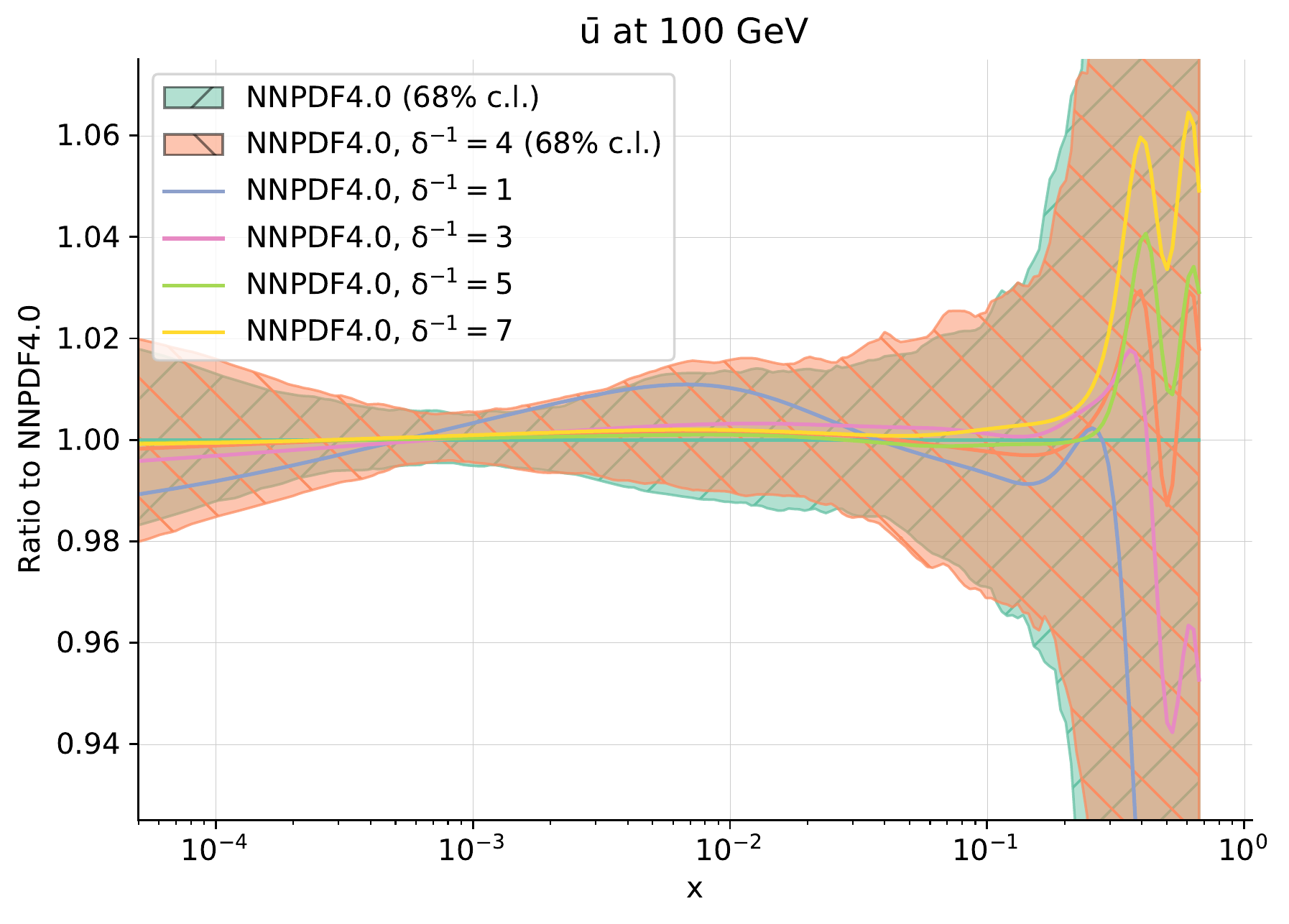}\\
  \includegraphics[width=0.48\textwidth]{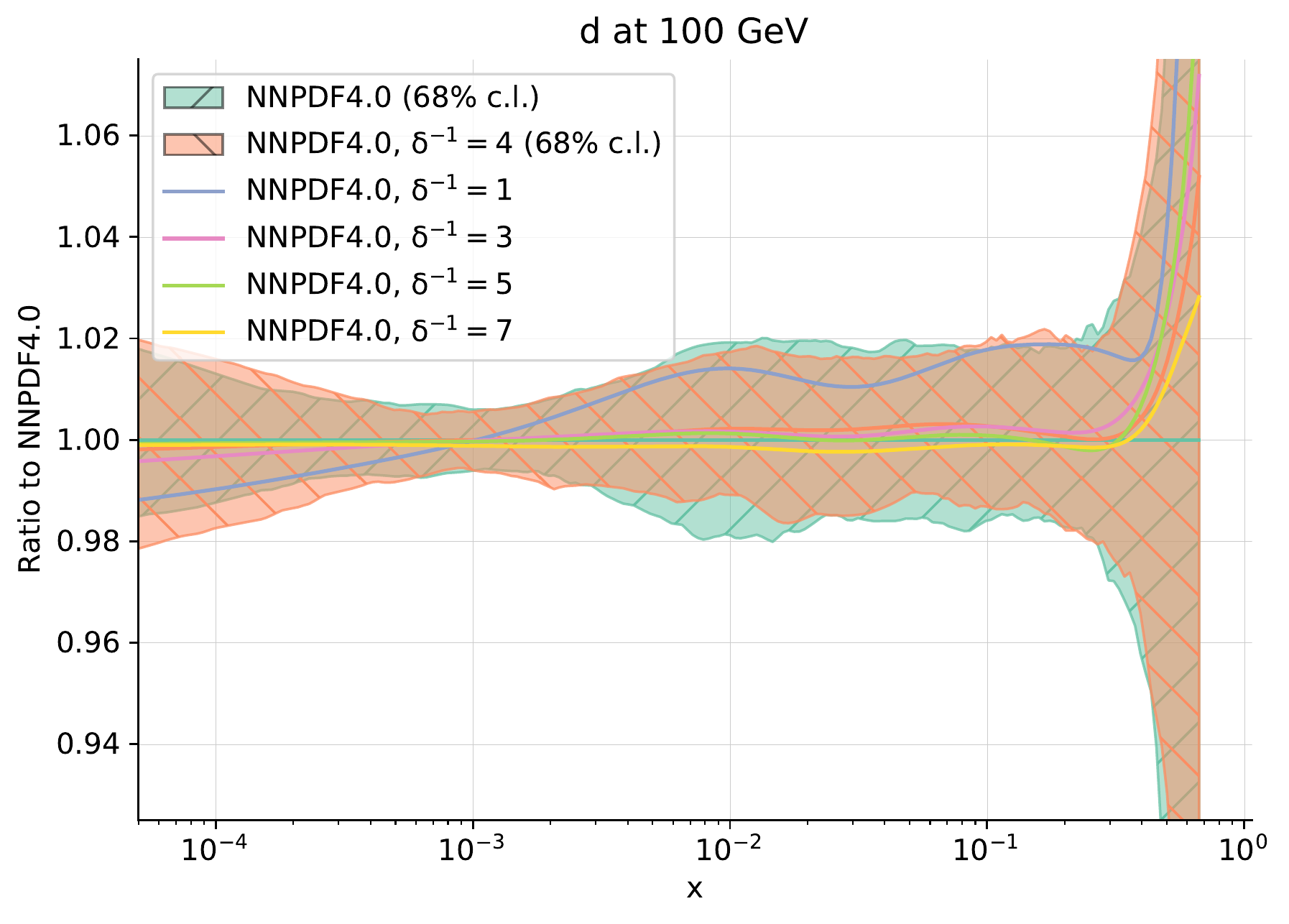}
  \includegraphics[width=0.48\textwidth]{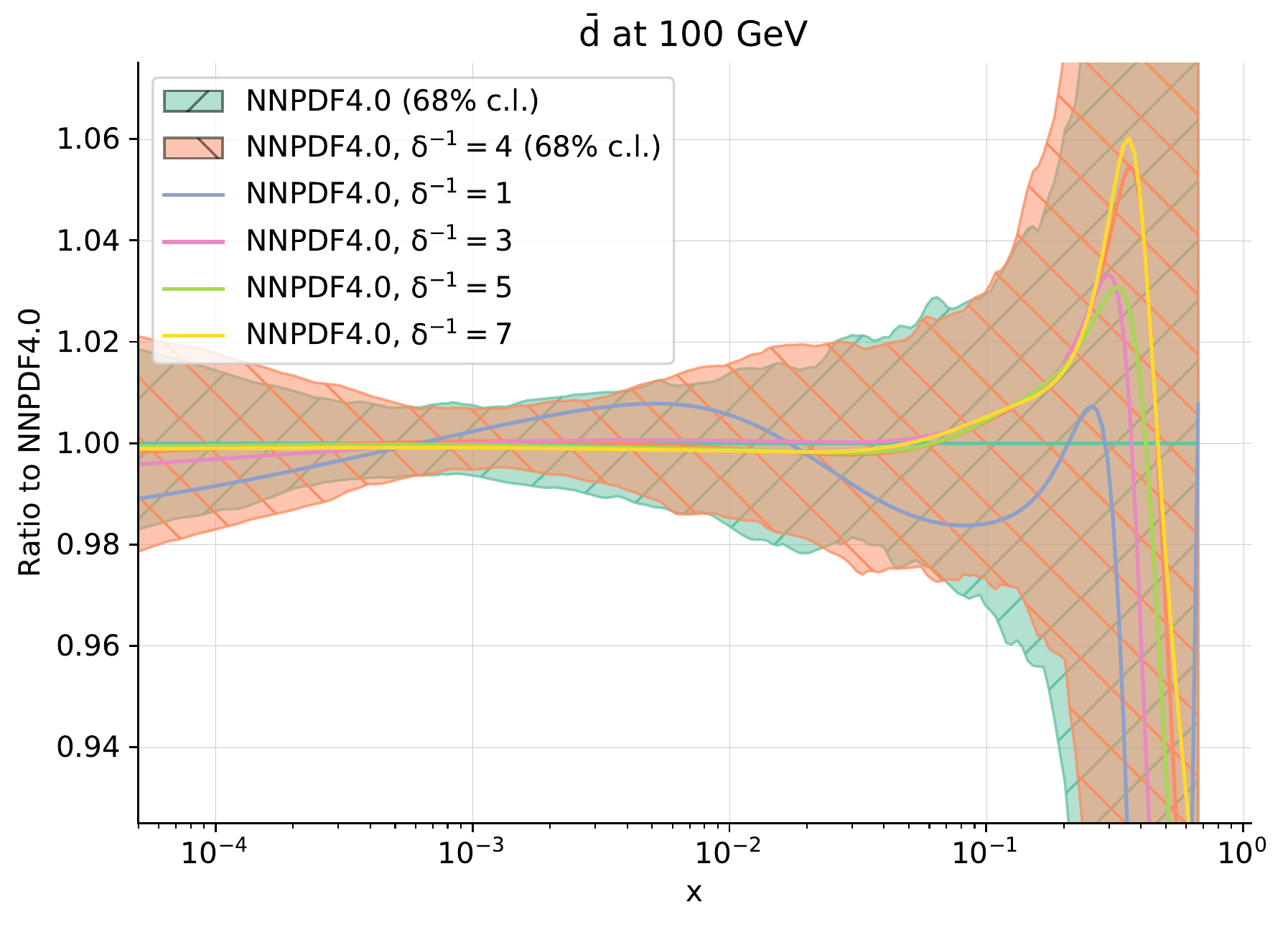}\\
  \includegraphics[width=0.48\textwidth]{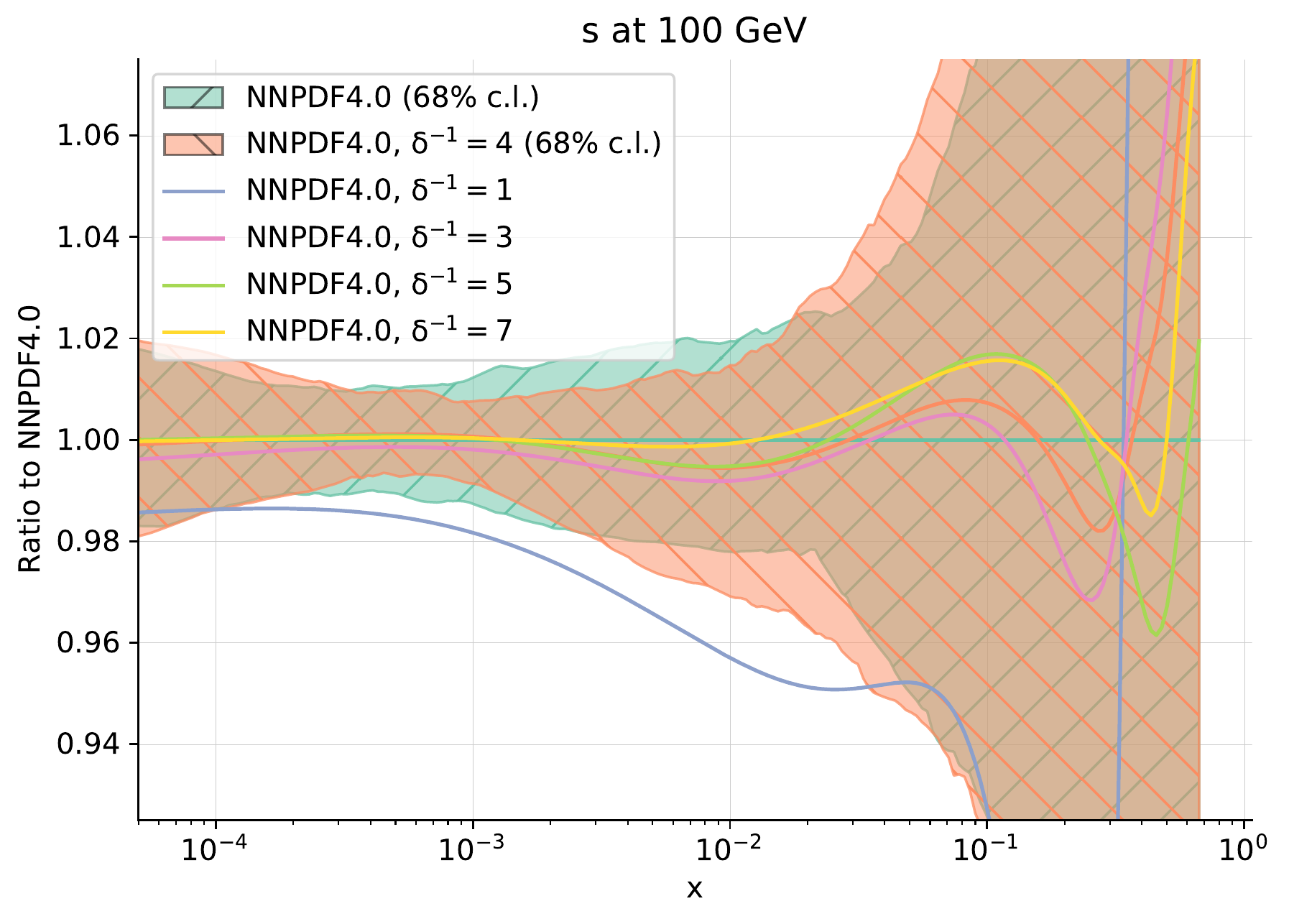}
  \includegraphics[width=0.48\textwidth]{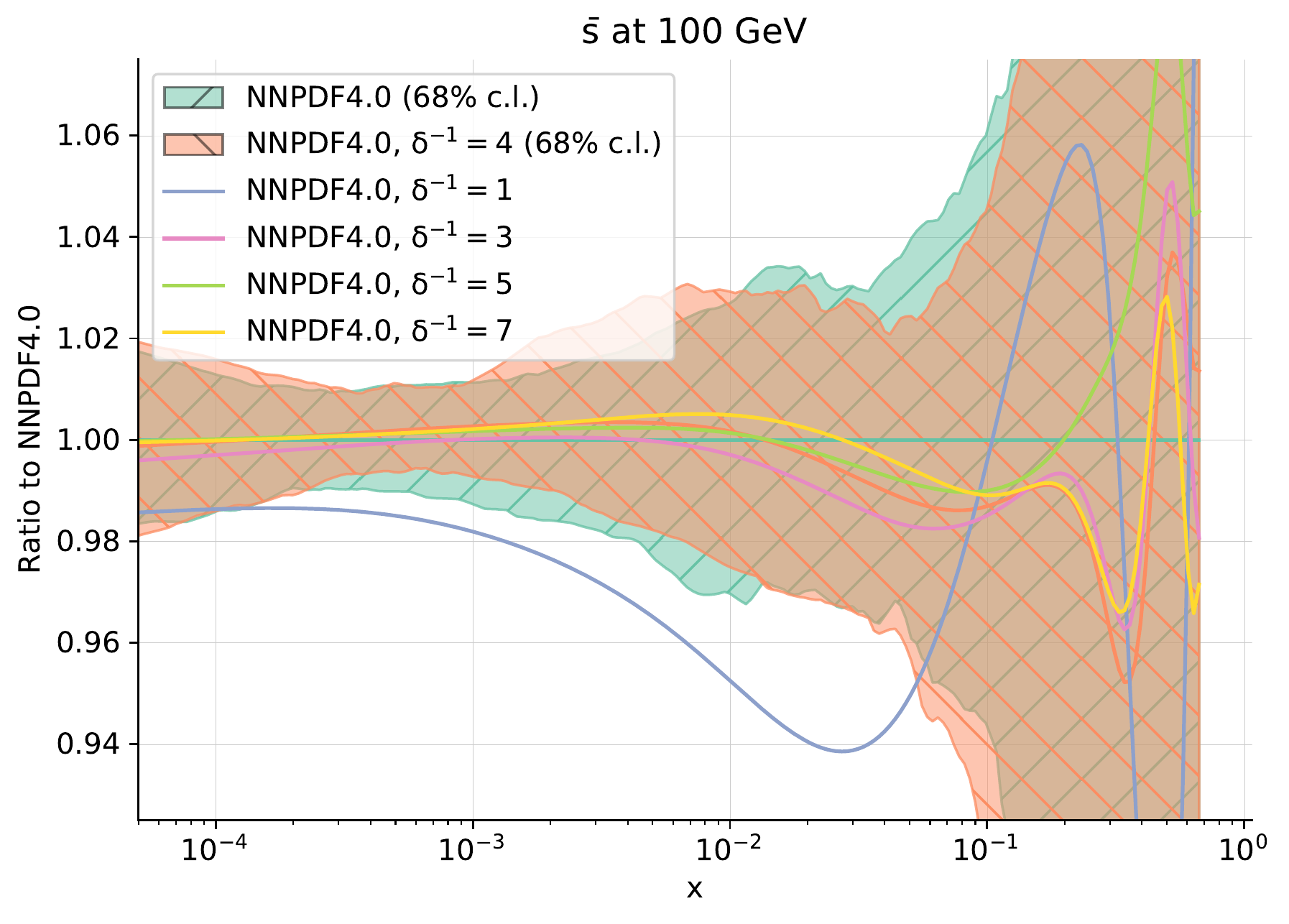}\\
  \includegraphics[width=0.48\textwidth]{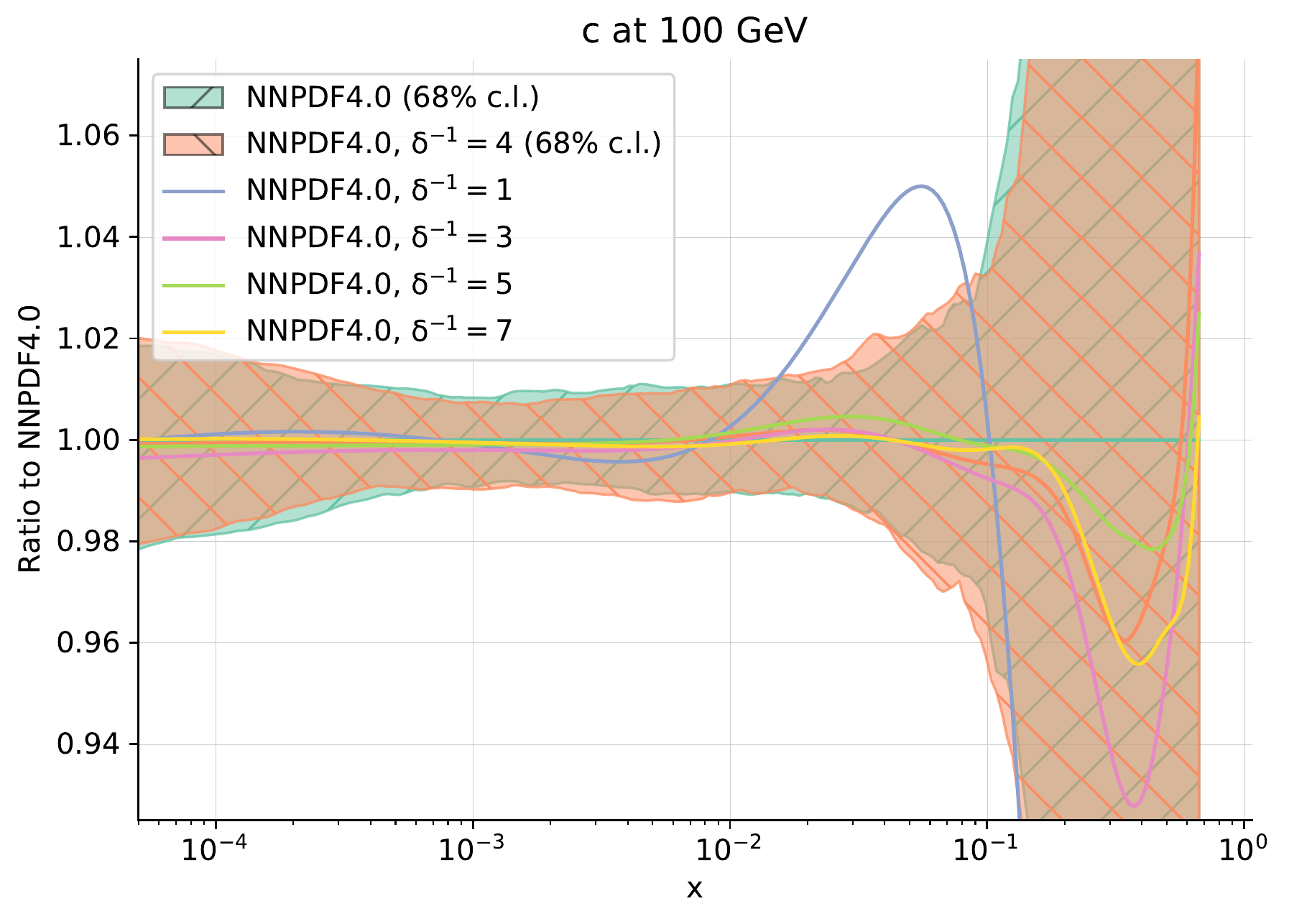}
  \includegraphics[width=0.48\textwidth]{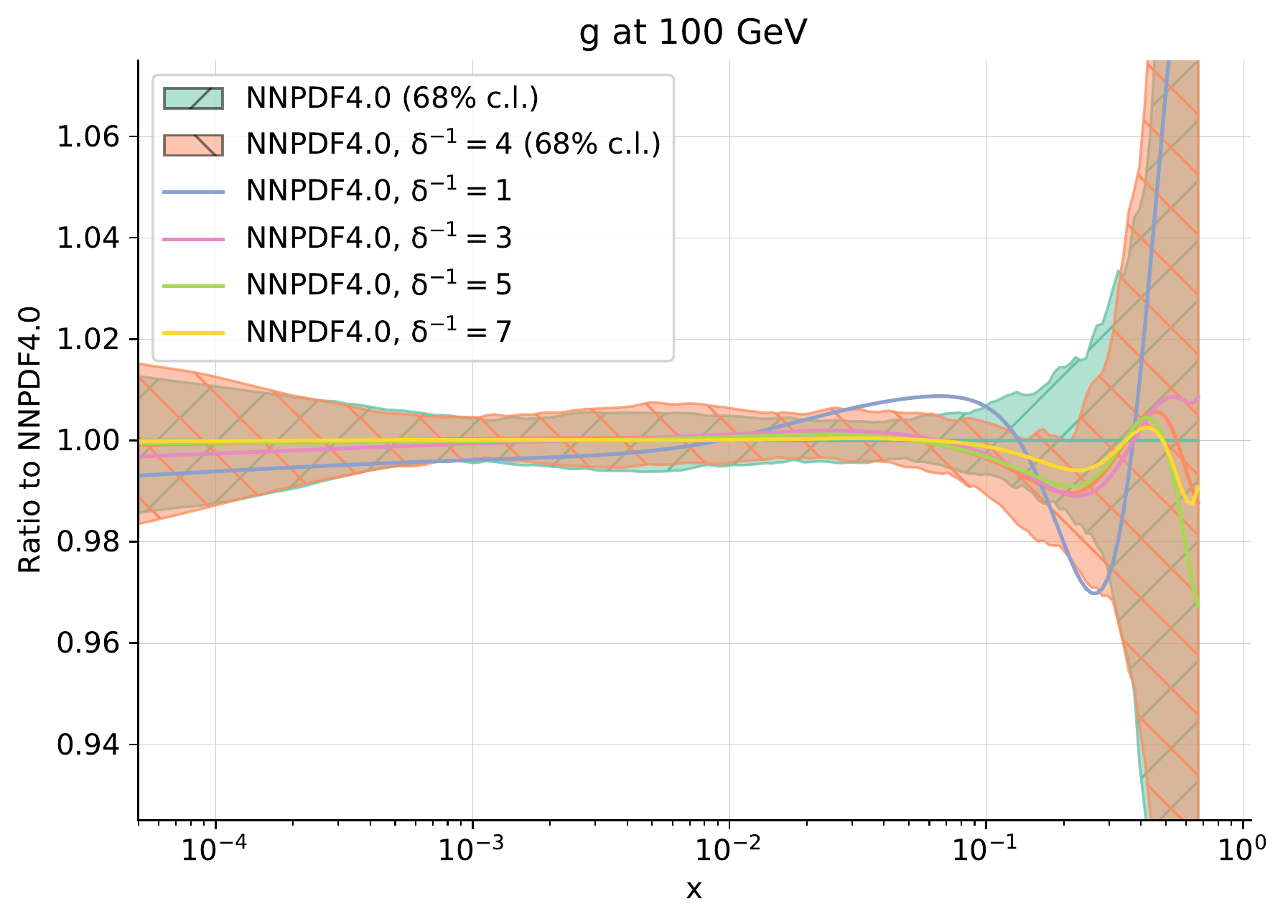}\\
  \vspace{0.5cm}
  \caption{The PDFs obtained by fitting the NNPDF4.0 data set after
    regularisation with different values of the parameter
    $\delta^{-1}=1,3,4,5,7$.
    From top to bottom, left to right, we show the up, anti-up, down,
    anti-down, strange, anti-strange, charm and gluon PDFs
    at a scale $Q=100$~GeV. PDFs are compared to the NNPDF4.0 baseline parton
    set, and normalised to its central value. For $\delta^{-1}=1,3,5,7$ we
    display only the central value. Otherwise uncertainties correspond to 68\%
    confidence levels. All PDF fits are accurate to NNLO.}
  \label{fig:PDF_fits}
\end{figure}

In Fig.~\ref{fig:PDF_fits} we then display the resulting PDFs, specifically the
up, anti-up, down, anti-down, strange, anti-strange, charm and gluon PDFs at a
scale $Q=100$~GeV. PDFs are compared to the NNPDF4.0 NNLO baseline parton
set, and are normalised to its central value. For $\delta^{-1}=1,3,5,7$ we
display only the central value. Otherwise uncertainties correspond to 68\%
confidence levels.

A joint inspection of Table~\ref{tab:chi2} and of Fig.~\ref{fig:PDF_fits}
reveals some interesting features. We first observe that, as expected, the
regularisation procedure has a significant effect on the $\chi^2$. A general
decrease of its value is observed in comparison to NNPDF4.0, by an amount that
increases with the increase in the amount of regularisation (that is, with the
decrease of the value of $\delta^{-1}$). For the largest value $\delta^{-1}=7$,
no statistically significant differences are seen with respect to NNPDF4.0,
neither in the value of the $\chi^2$ per data point nor in PDFs. Conversely,
for small values of $\delta^{-1}$, $\delta^{-1}=1$ and $\delta^{-1}=2$, the total
$\chi^2$ per data point drops from $1.16$ to $0.58$ and $0.97$, respectively.
These variations correspond to a $28\sigma$ and a $9\sigma$ fluctuation in
units of the $\chi^2$ standard deviation, which obviously denote an excessive
regularisation of the NNPDF4.0 data set. As noted at the end of
Sect.~\ref{sec:regularisation}, such an excessive regularisation may also
arise from neglecting terms of $\mathcal{O}(\delta^2)$ in
Eq.~\eqref{eq:deltachi2bound0}.

The PDFs obtained in the fit with $\delta^{-1}=1$ (and similarly in the fit with
$\delta^{-1}=2$, which is not displayed in Fig.~\ref{fig:PDF_fits}) are indeed
consistently distorted in comparison to NNPDF4.0. The central value of the
former fluctuates, in units of the NNPDF4.0 PDF uncertainty around the central
value of the latter, by about one standard deviation for the up, anti-up, down
and anti-down PDFs, and slightly more for the strange, anti-strange, charm and
gluon PDF. In this respect, it is worth noting that the strange and gluon PDFs
are sensitive, respectively, to the ATLAS $W,Z$~7~TeV~CC~\cite{ATLAS:2016nqi}
and ATLAS dijets~R=0.6~7~TeV~\cite{ATLAS:2013jmu} data sets: these have some
of the largest values of $Z$ and display the largest reduction of $\chi^2$
upon regularisation.

The outer cases corresponding to $\delta^{-1}=1,2,7$ are therefore to be
interpreted as a validation of the regularisation procedure, which behaves as
expected.  The fits corresponding to $\delta^{-1}=3$, $\delta^{-1}=4$ and
$\delta^{-1}=5$ are instead more interesting. Variations of the $\chi^2$ with
respect to NNPDF4.0 correspond, respectively, to a $3.8\sigma$, $2.4\sigma$ and
$1.4\sigma$ fluctuation in units of the $\chi^2$ standard deviation.
Interestingly, the difference between the expected $\chi^2/N_{\rm dat}=1$ and
the $\chi^2$ obtained in the fits corresponding to $\delta^{-1}=3,4,5$ amounts,
respectively, to $3.3\sigma$, $5.3\sigma$ and $6.2\sigma$ in units of the
$\chi^2$ standard deviation.  This is a significant reduction in
comparison to $7.7\sigma$ of the default NNPDF4.0 determination.

Such an improvement in the $\chi^2$ statistic is accompanied by remarkably limited PDF
variations if one compares the fits with $\delta^{-1}=3,4,5$ with NNPDF4.0. Central
values fluctuate by a small fraction of the NNPDF4.0 PDF uncertainty, except
for the gluon PDF, which varies by up to half of the NNPDF4.0 uncertainty
around $x\sim 0.3$; PDF uncertainties are almost unaffected. Remarkably, all
these variations are much smaller than those due to variations of the data set
itself (see Sect.~7 in~\cite{NNPDF:2021njg}). 

\begin{figure}[!t]
  \centering
  \includegraphics[width=0.48\textwidth]{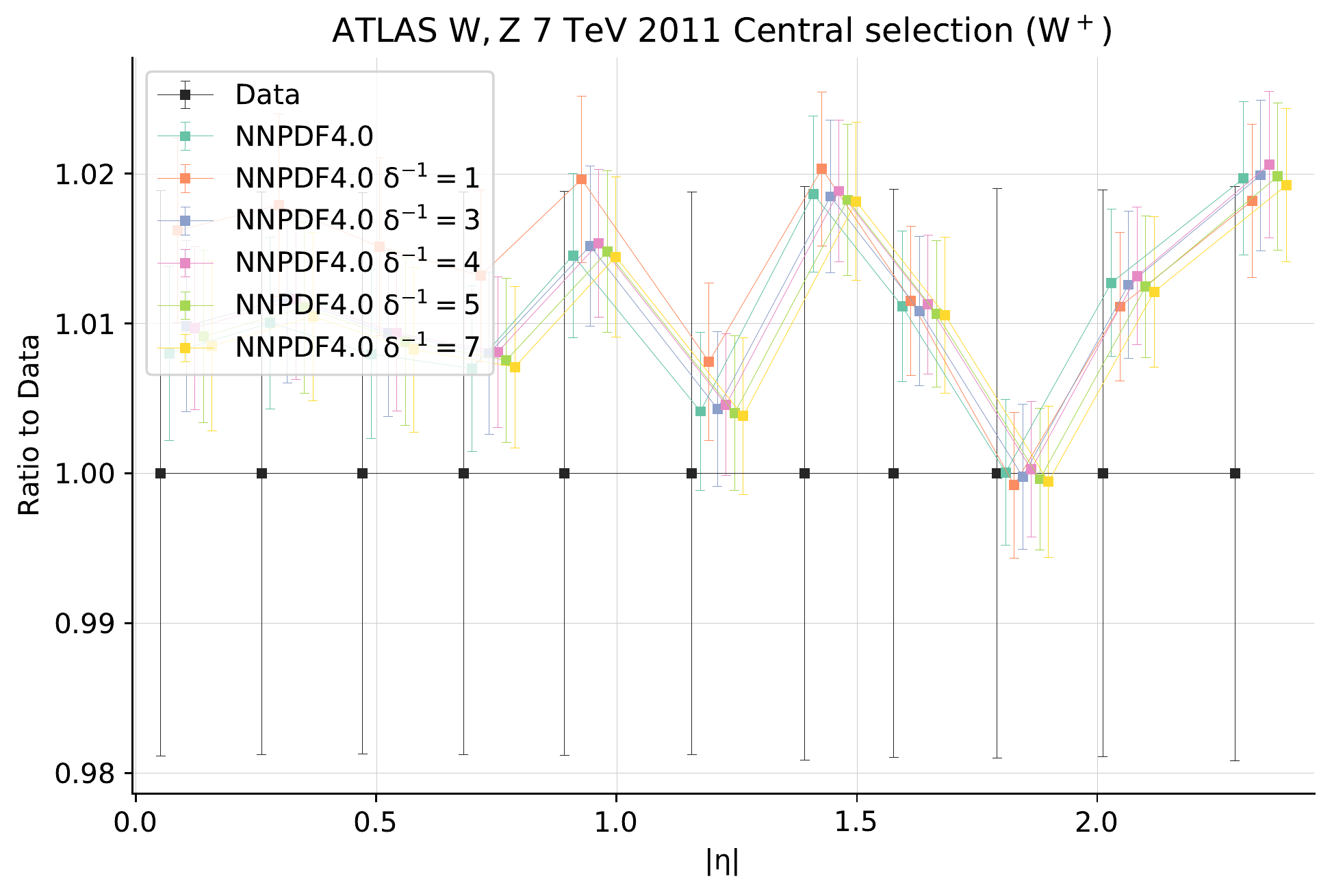}
  \includegraphics[width=0.48\textwidth]{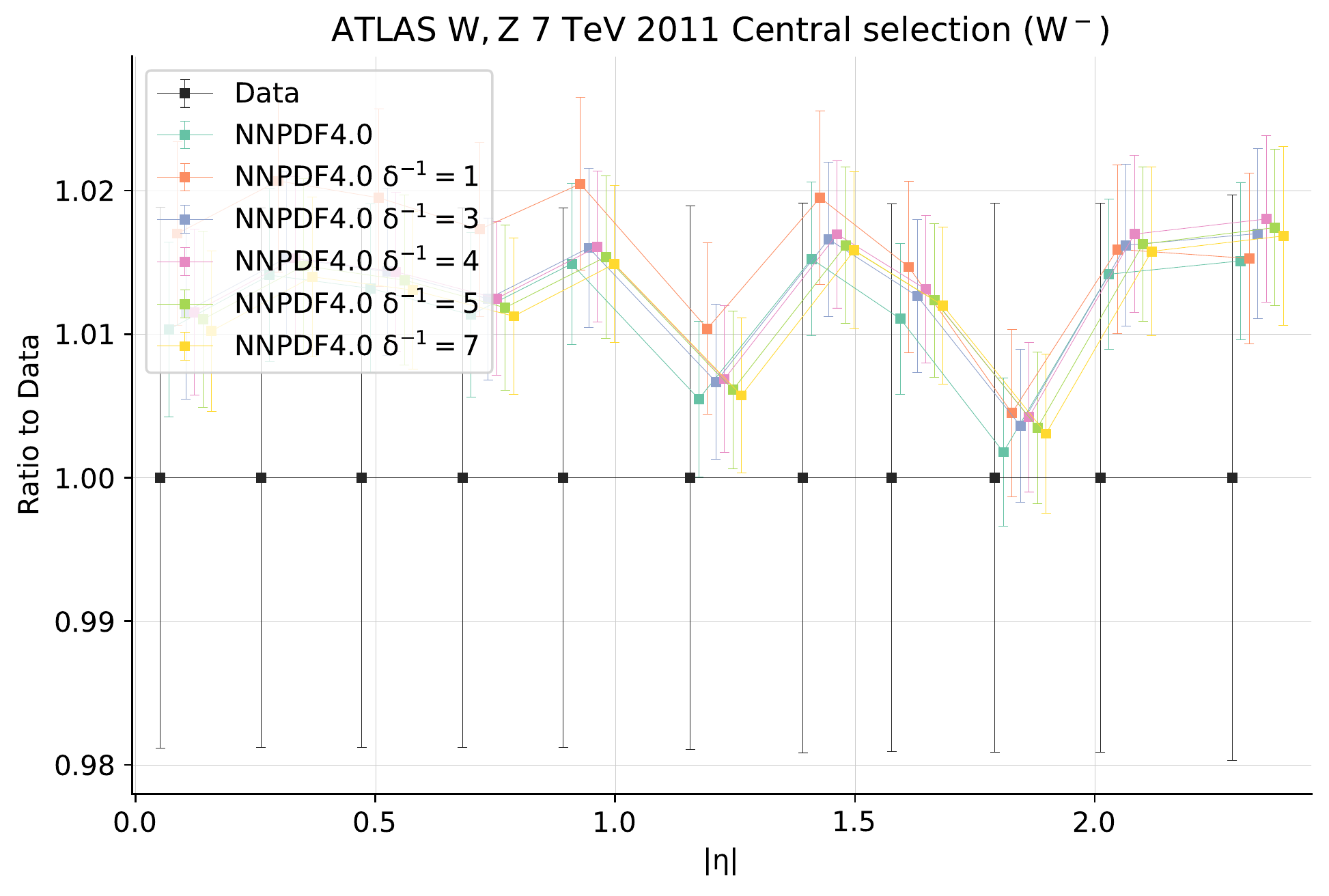}\\
  \includegraphics[width=0.48\textwidth]{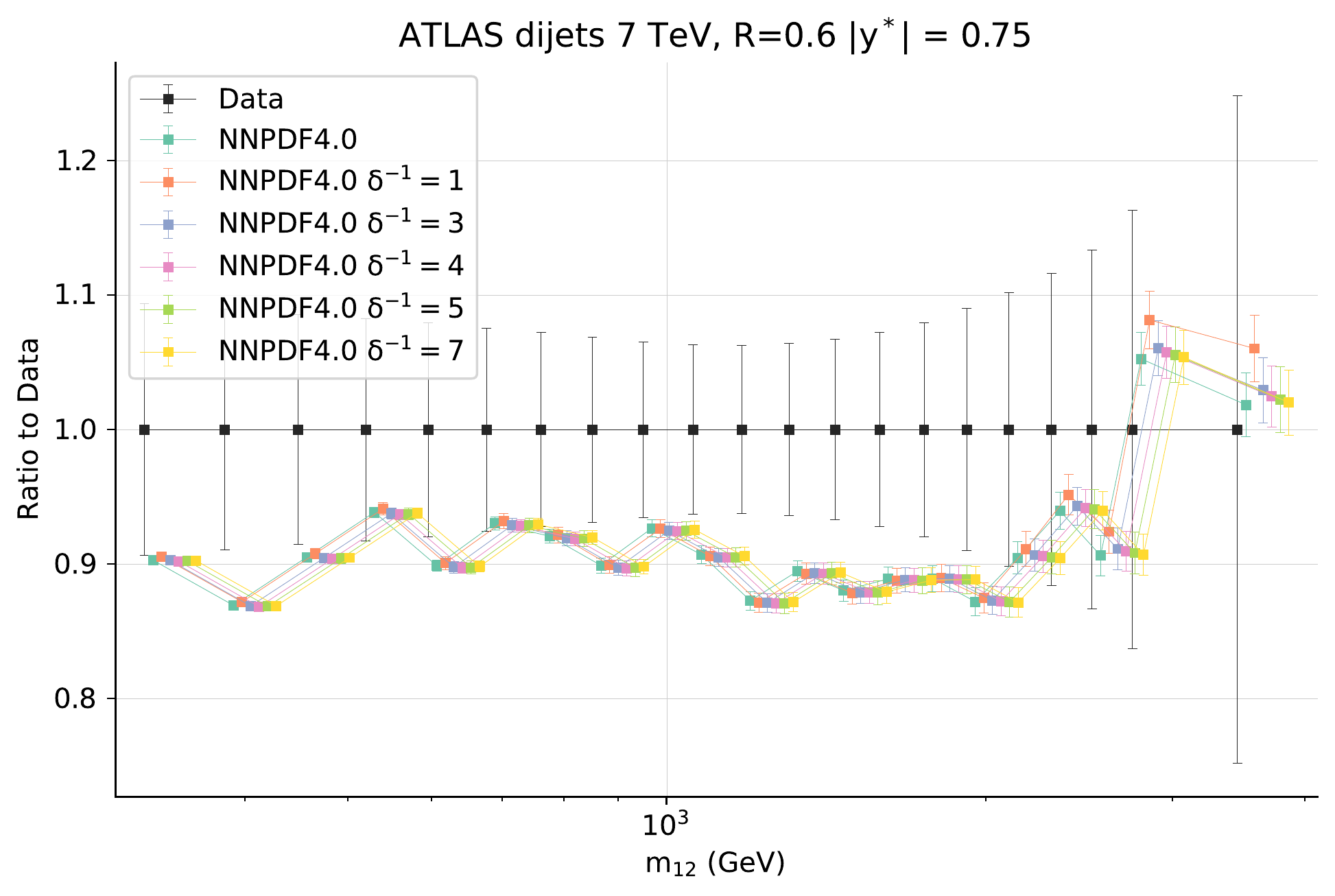}
  \includegraphics[width=0.48\textwidth]{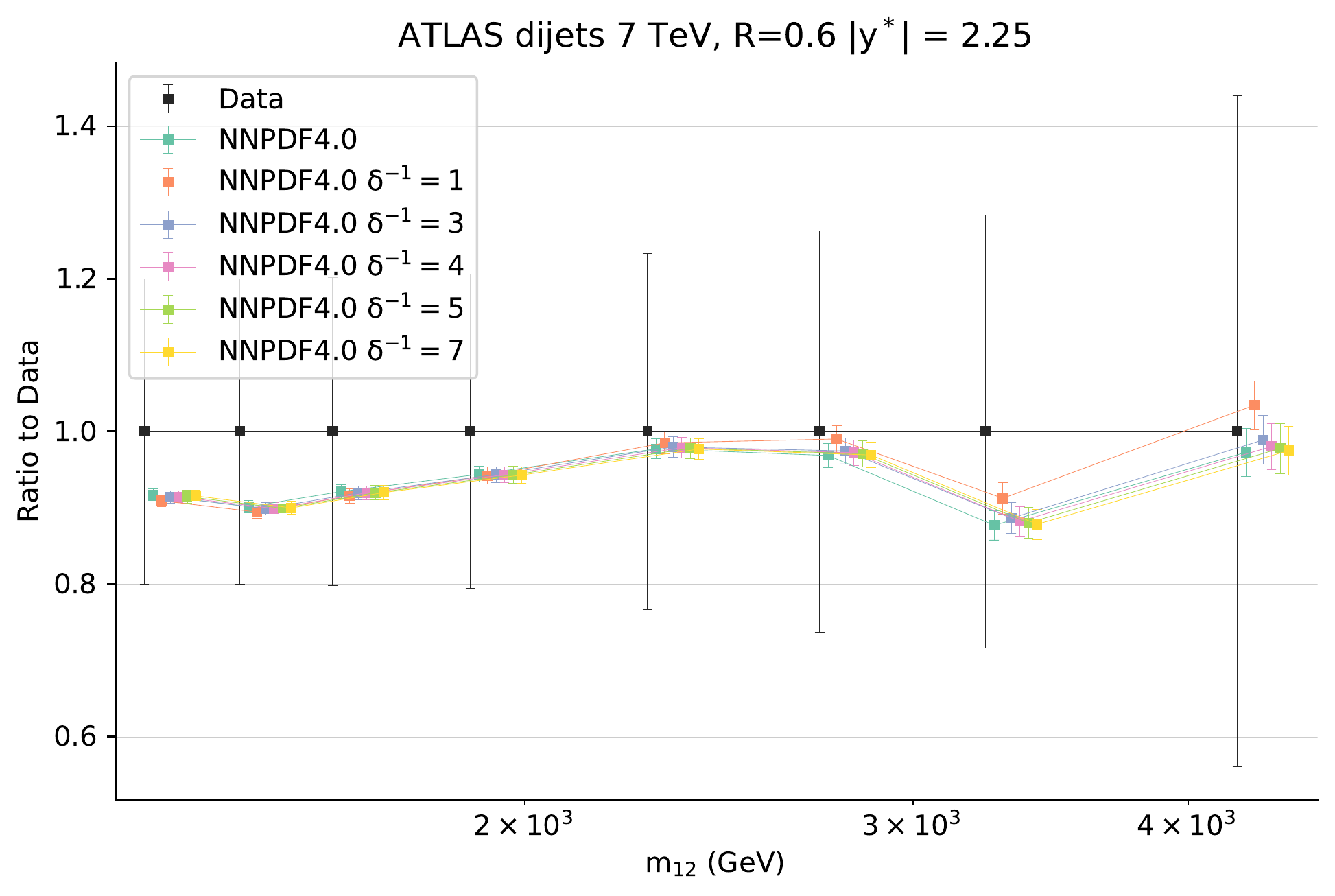}\\
  \vspace{0.5cm}
  \caption{Data--theory comparison for the $W^\pm$ subset of the ATLAS
    $W,Z$~7~TeV~CC measurement~\cite{ATLAS:2016nqi}, as a function of the
    absolute lepton rapidity $\eta$ (top), and for two bins in the absolute
    rapidity difference between the two leading jets $|y^*|$ of the ATLAS
    dijets~R=0.6~7~TeV measurement~\cite{ATLAS:2013jmu}, as a function of the
    di-jet invariant mass $m_{12}$. Theoretical predictions are obtained with
    the NNPDF4.0 baseline parton set and with the PDFs obtained by fitting the
    NNPDF4.0 data set regularised with $\delta^{-1}=1,3,4,5,7$. They are all
    accurate to NNLO in the strong coupling, both in the PDFs and in the matrix
    elements. Results are shown as ratios to the experimental central value,
    with one-sigma PDF and experimental uncertainties. The latter is the sum in
    quadrature of the statistical and of all systematic uncertainties.}
  \label{fig:data_theory}
\end{figure}

The fact that PDFs do not vary significantly in the fits to the regularised
data set with $\delta^{-1}=3,4,5$ is further displayed in Fig.~\ref{fig:data_theory},
where we show a data--theory comparison for some selected bins of the
ATLAS $W,Z$~7~TeV~CC~\cite{ATLAS:2016nqi} and
dijets~R=0.6~7~TeV~\cite{ATLAS:2013jmu} measurements. Specifically, we show the
$W^+$ and $W^-$ subsets, as a function of the absolute value of the lepton
rapidity $\eta$, for the former, and two bins in the absolute rapidity
difference between the two leading jets $|y^*|$ as a function of the di-jet
invariant mass $m_{12}$, for the latter. Theoretical predictions are obtained
with the NNPDF4.0 baseline parton set and with the PDFs obtained by fitting the
NNPDF4.0 data set regularised with $\delta^{-1}=1,3,4,5,7$. They are all accurate
to NNLO in the strong coupling, both in the PDFs and in the matrix elements.
Results are shown as ratios to the experimental central value, with one-sigma
experimental and PDF uncertainties. The experimental uncertainty is the sum in
quadrature of the statistical and of all systematic uncertainties.

As noted in Sect.~\ref{subsec:characterisation}, the data sets displayed in
Fig.~\ref{fig:data_theory} are those with large values of $Z$ and $\chi^2$, and
for which the regularisation procedure introduces some of the largest
differences in the variances and correlations of the data, see
Table~\ref{tab:differences} and Fig.~\ref{fig:covmatrices}. In spite of this,
only small differences are
observed between predictions obtained with NNPDF4.0 and any of the regularised
fits with $\delta^{-1}=1,3,4,5,7$; slightly larger fluctuations are observed in
the fit with a large amount of regularisation ($\delta^{-1}=1$), albeit only for
the data points at central rapidity, for ATLAS $W,Z$~7~TeV~CC, or at large
invariant mass, for ATLAS dijets~R=0.6~7~TeV.

We therefore conclude that the PDFs obtained from any of the regularised fits
with $\delta^{-1}=3,4,5$ represent the same underlying truth as the NNPDF4.0
parton set. They however lead to a $\chi^2$ that is better than the NNPDF4.0 one
by up to $4\sigma$, in units of the $\chi^2$ standard deviation, and that is
only about $3\sigma$ away from the expectation of unit $\chi^2$ (instead of
about $8\sigma$). In other words, the nominal $\chi^2$ determined
in~\cite{NNPDF:2021njg} is likely to be spuriously inflated by inaccuracies in
the estimation of the experimental correlations in the LHC data. Further
discrimination among the equally good values $\delta^{-1}=3,4,5$ can be made
on the basis of how big the changes to the covariance matrices are
in relation to the precision at which they are estimated. Since the precision is
unknown, this entails a degree of subjectivity. We deem that the values of
$\Delta\sigma_r < 5\%$ and $|\Delta\rho| < 0.05$ implied by $\delta^{-1}=4$
suggest it safe to assume that the resulting regularised covariance matrices are
compatible with the original ones within the precision at which they were
determined, while ensuring stability against possibly bigger inaccuracies in the
correlations. Therefore, the fit with $\delta^{-1}=4$ will be used as reference
in the remainder of this paper.

\subsection{Correlating and decorrelating experimental uncertainties
  with more information}
\label{subsubsec:models}

As we have mentioned in Sect.~\ref{subsec:regularisation_procedure}, the
correlation models provided with the measurements have to be preferred to our
regularisation procedure whenever these are available, and if they result in a
stable covariance matrix. For example, the
correlation model recommended in~\cite{ATLAS:2017kux} for the analysis of the
ATLAS jets R=0.6 8~TeV measurement is used by default in the NNPDF4.0
determination~\cite{NNPDF:2021njg} and in all the fits presented in
Sect.~\ref{subsec:fits}. It is therefore not surprising that the regularisation
procedure has almost no impact on the $\chi^2$ of this specific data set.

Correlation models, which follow from a careful experimental analysis of all of
the sources of systematic uncertainties and of their correlations, are however
not always available. Sometimes they become available only long after the
measurement is published, and sometimes a clear recommendation for their usage
is not provided. In order to remedy this lack of information, some guesswork
is carried out to identify the systematic uncertainties whose nominal
correlations are likely to be too strong. For instance, two of
these~\cite{Harland-Lang:2017ytb,Bailey:2019yze} have targeted, respectively,
the ATLAS 7~TeV single-inclusive jet measurement~\cite{ATLAS:2014riz}
and the 8~TeV top-pair lepton+jet measurement~\cite{ATLAS:2015lsn}. They were
performed in the framework of the MMHT2014 global
analysis~\cite{Harland-Lang:2014zoa} by inspecting the nuisance parameters
associated to each systematic uncertainty in the $\chi^2$. Similar
studies~\cite{Nocera:2017zge,Amoroso:2020lgh}, targeting the same measurements
and based on complete decorrelation of certain systematic uncertainties, were
also carried out in the framework on the NNPDF3.1 global
analysis~\cite{NNPDF:2017mvq}. Sometimes these analyses have been used to
inform and/or validate the experimental correlation models. In this respect,
our regularisation procedure can be utilised in the same spirit, with the
advantage that it is more general and requires less information than the
aforementioned analyses.

Here we investigate how the regularisation procedure performs in comparison to
the correlation models provided with the measurement in the few cases in which
these are available. We consider two cases. The first case concerns the ATLAS
dijets~R=0.6~7~TeV~\cite{ATLAS:2013jmu} measurement, for which a STRONG and a
WEAK (de-)correlation models are provided on top of the nominal correlation
model used in NNDPF4.0 and in all the fits of Sect.~\ref{subsec:fits}. None of
these models are clearly recommended in~\cite{ATLAS:2013jmu}, hence why they
have not been previously considered. The second case concerns three ATLAS
8~TeV measurements, namely the $W^\pm$+jet~\cite{ATLAS:2017irc}, the
$t\bar{t}~\ell$+jets~\cite{ATLAS:2015lsn}, and the single-inclusive jets
R=0.6~\cite{ATLAS:2017kux} measurements. Details on how to correlate or
decorrelate systematic uncertainties between bins within and across these
measurements have been provided only very recently~\cite{ATLAS:2021vod}. This
is the reason why they have not been previously considered. We will refer to
this correlation model with the label ATLAS henceforth.

We then perform four fits, all based on the experimental, theoretical, and
methodological input that enters the default NNPDF4.0 parton set, by considering
these correlation models. The first two fits are performed using, respectively,
the STRONG and WEAK correlation models for the ATLAS dijets~R=0.6~7~TeV
measurement. Experimental correlations for all of the other data sets are
otherwise as in NNPDF4.0. The third fit is performed using the ATLAS
correlation model for all the concerned ATLAS 8~TeV measurements. This
correlation model was not completely utilised in NNPDF4.0 (in particular for
what concerns correlations between pairs of points belonging to different data
sets). It also does not enter the two aforementioned fits. The fourth fit
is performed by combining the WEAK and ATLAS correlation models at the same
time.

In Table~\ref{tab:chi2_models} we display the value of the $\chi^2$ per data
point, $\chi^2/N_{\rm dat}$, for each of these fits and compare it to that of the
NNLO NNPDF4.0 default fit, and of the fit obtained by regularising the
NNPDF4.0 data set with $\delta^{-1}=4$. For conciseness, we aggregate the data
sets into one of the following classes: deep-inelastic scattering,
fixed-target Drell--Yan, Tevatron Drell--Yan, ATLAS, CMS, and LHCb. For ATLAS,
we also indicate the individual $\chi^2$ of the data sets affected by the
correlation models. The corresponding number of data points, $N_{\rm dat}$,
is also indicated.

\begin{table}[!t]
  \scriptsize
  \centering
  \renewcommand{\arraystretch}{1.4}
  \input{tables/tab-chi2_models.tex}\\
  \vspace{0.5cm}
  \caption{The number of data points, $N_{\rm dat}$, and the $\chi^2$ per data
    point, $\chi^2/N_{\rm dat}$, for the NNPDF4.0 NNLO baseline fit, for each of
    the fits performed with a different correlation model (see text for
    details), and for the fit to the NNPDF4.0 data set regularised with
    $\delta^{-1}=4$.}
  \label{tab:chi2_models}
\end{table}

\begin{figure}[!t]
  \centering
  \includegraphics[width=0.48\textwidth]{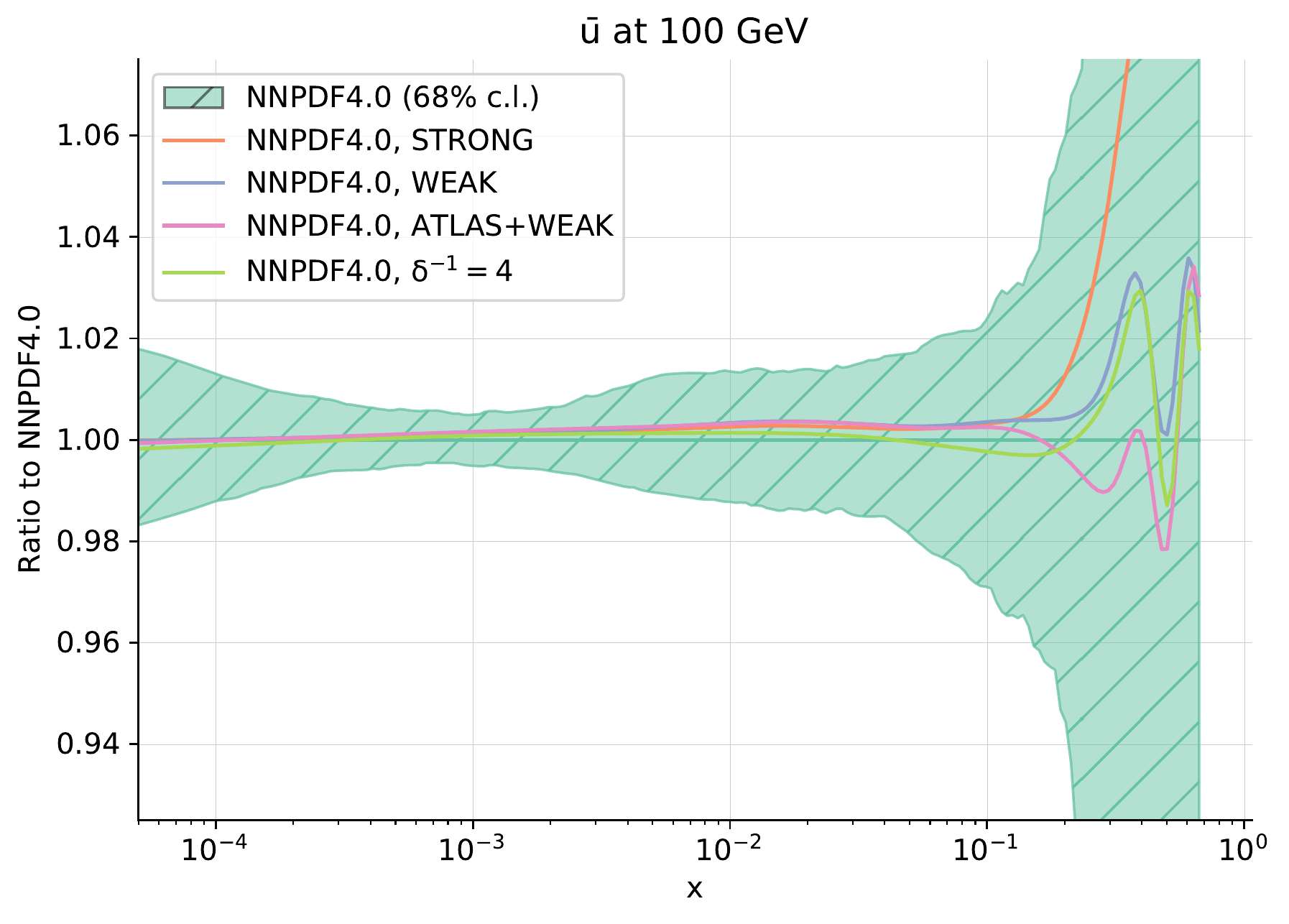}
  \includegraphics[width=0.48\textwidth]{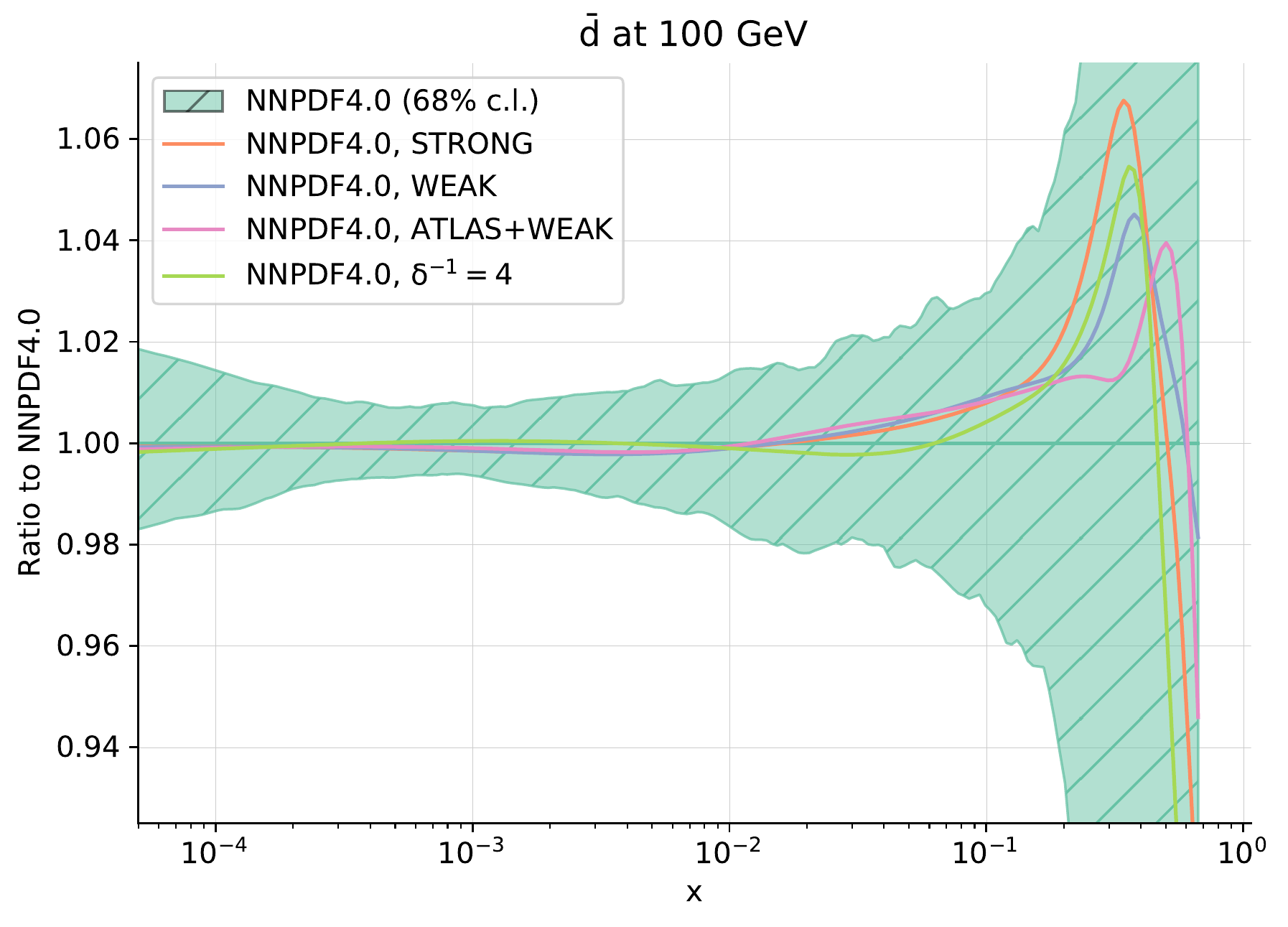}\\
  \includegraphics[width=0.48\textwidth]{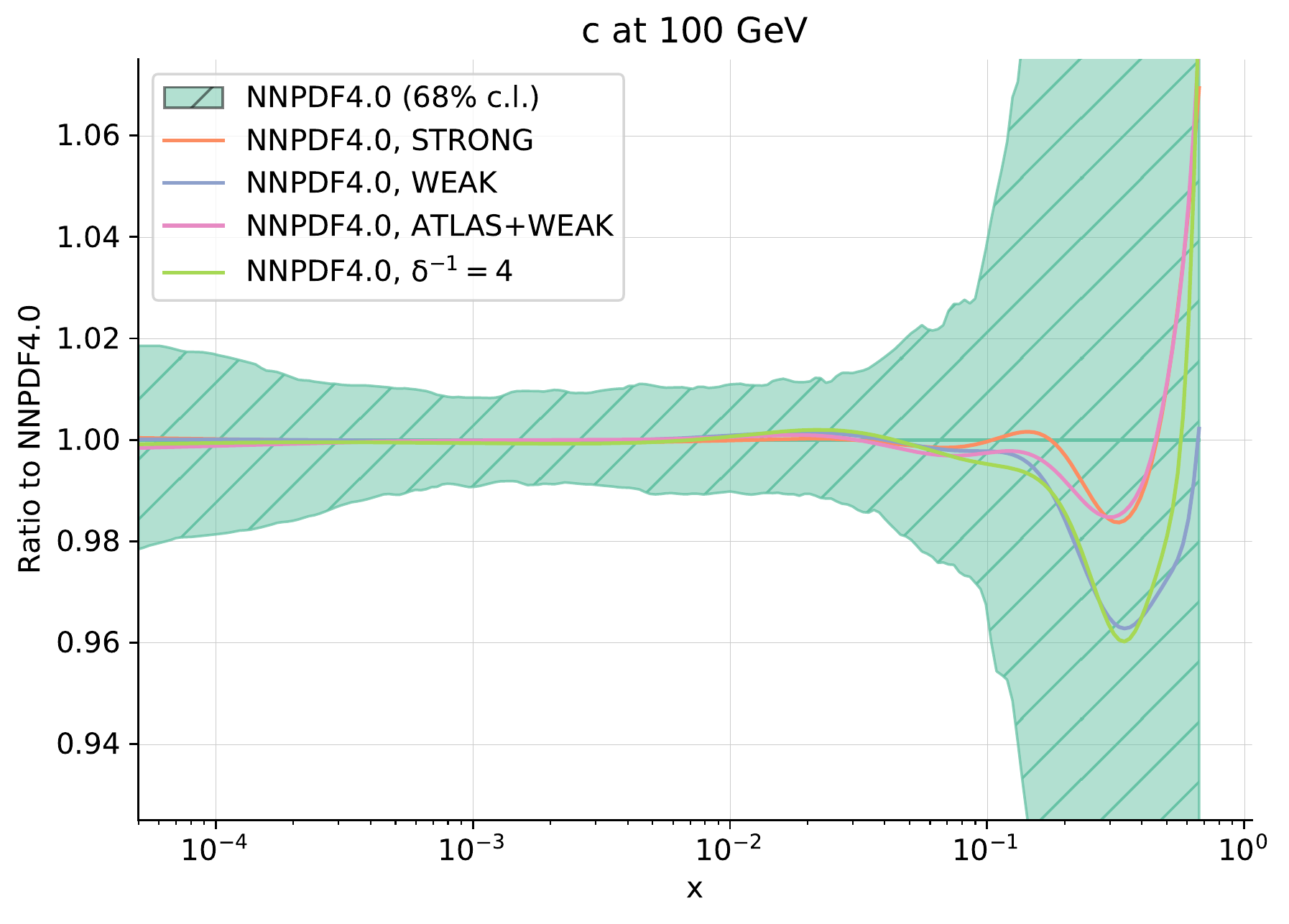}
  \includegraphics[width=0.48\textwidth]{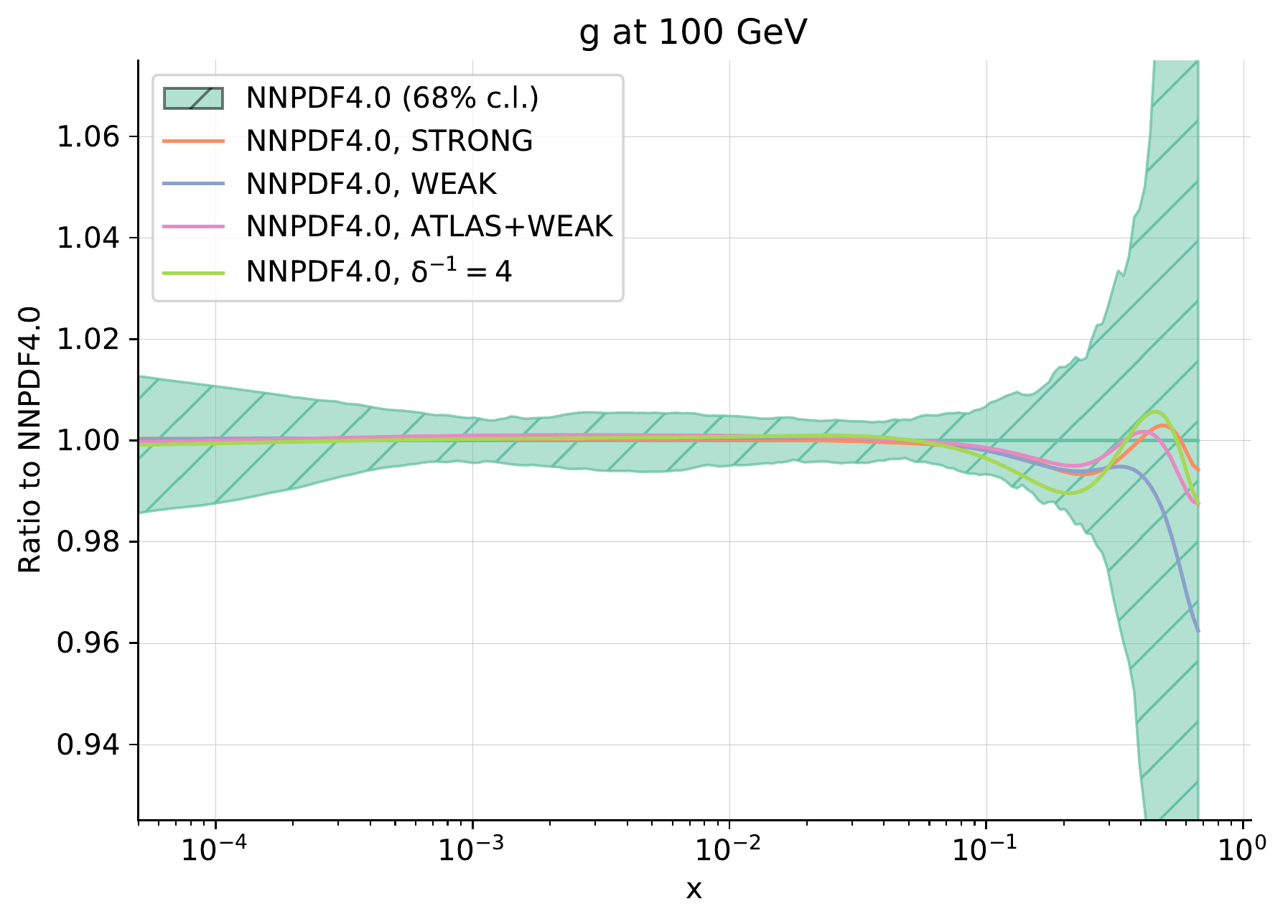}\\
  \vspace{0.5cm}
  \caption{The PDFs obtained by fitting the NNPDF4.0 data set with correlation
    models provided by the experiment for a subset of measurements (see text
    for details). From top to bottom, left to right, we show the anti-up,
    anti-down, charm and gluon PDFs at a scale $Q=100$~GeV. PDFs are compared
    to the NNPDF4.0 baseline parton set, and normalised to its central value.
    Also shown are the PDFs obtained in a fit to the NNPDF4.0 data set
    regularised with $\delta^{-1}=4$. For all PDFs but the NNPDF4.0 baseline, we
    show only the central value. Otherwise the uncertainty corresponds to the
    68\% confidence interval. All PDFs are accurate to NNLO.}
  \label{fig:PDFs_corr_model}
\end{figure}

In Fig.~\ref{fig:PDFs_corr_model} we show the resulting PDFs, specifically the
anti-up, anti-down, charm and gluon PDFs at a scale $Q=100$~GeV. PDFs are
compared to the NNPDF4.0 NNLO baseline parton set, and to the PDFs obtained
by regularising the NNPDF4.0 data set with $\delta^{-1}=4$. All the curves are
normalised to the NNPDF4.0 central value. For all PDFs but the
NNPDF4.0 NNLO baseline, we show only the central value. Otherwise the
uncertainty corresponds to the 68\% confidence interval.

A joint inspection of Table~\ref{tab:chi2_models} and
Fig.~\ref{fig:PDFs_corr_model} reveals two features. First, the fit quality,
as quantified by the value of the $\chi^2$ per data point, does not change
upon variation of the available correlation models, either for the data sets
affected by the model, or for the other data sets. This behaviour contrasts
with the larger variations seen upon refitting a regularised data set, even
when the amount of regularisation is fairly limited, see Table~\ref{tab:chi2}. 
Second, the shifts of PDF central values induced by a given correlation
model, however modest they turn out to be, are generally very close to the
shifts induced by the regularisation of the data set (specifically with
$\delta^{-1}=4$).
This is apparent for the WEAK correlation model, whose only feature is to
partially decorrelate certain uncertainties in one of the data sets with the
largest condition number, ATLAS dijets~R=0.6~7~TeV~\cite{ATLAS:2013jmu}. In
this respect, this correlation model is relatively close to what the
regularisation procedure achieves. For other correlation models, qualitatively
similar shifts are also seen, although with some quantitative differences.

The fact that our regularisation procedure captures the same qualitative
shifts on PDF central values as experimental correlation models, once one or
the others are used to determine the PDFs, is suggestive. Whether this is a
coincidental feature, limited to the correlation models considered, or a
more general one, could only be investigated if additional correlation models
become available to be tested. In general, it is reasonable that a correlation
model altering the original covariance matrix as little as possible while
improving the stability of the $\chi^2$ leads to results similar to those
obtained with the regularisation procedure. That being said, the shift in
central value remains so small that it would be hard to make any conclusions
based on its statistical significance.

On the other hand, the fact that using correlation models does not lead to a
better $\chi^2$ is a consequence of the fact that the condition number of the
corresponding experimental covariance matrix is almost unaltered, as we have
explicitly checked. We therefore conclude that the available correlation models
are not enough to lead to a stable experimental covariance matrix and $\chi^2$.
In light of these considerations, we find that our regularisation procedure can
possibly be used as a useful diagnosis tool to inform and validate correlation
models not only at the level of PDF fits, but also at the level of the
corresponding experimental analyses.

%% file: tables/tab-differences.tex
\begin{tabularx}{\textwidth}{Xcccccccccccccc}
\toprule
&
&
& \multicolumn{2}{c}{$\delta^{-1}=1$}
& \multicolumn{2}{c}{$\delta^{-1}=2$}
& \multicolumn{2}{c}{$\delta^{-1}=3$}
& \multicolumn{2}{c}{$\delta^{-1}=4$}
& \multicolumn{2}{c}{$\delta^{-1}=5$}
& \multicolumn{2}{c}{$\delta^{-1}=7$}\\
Data set & Ref. & $Z$
& $\Delta\sigma_r$ & $|\Delta\rho|$
& $\Delta\sigma_r$ & $|\Delta\rho|$
& $\Delta\sigma_r$ & $|\Delta\rho|$
& $\Delta\sigma_r$ & $|\Delta\rho|$
& $\Delta\sigma_r$ & $|\Delta\rho|$
& $\Delta\sigma_r$ & $|\Delta\rho|$ \\
\midrule
ATLAS LM DY 7~TeV
& \cite{ATLAS:2014ape}
&  3.7
&  71.0 & 0.51 &  12.0 & 0.15 &  2.51 & 0.03 &      &      &      &      &      &       \\
ATLAS HM DY 7~TeV
& \cite{ATLAS:2013xny}
&  3.1
&  67.3 & 0.50 &  8.99 & 0.13 &  0.43 & 0.01 &      &      &      &      &      &       \\
ATLAS $W,Z$ 7~TeV
& \cite{ATLAS:2011qdp}
&  3.5
&  87.2 & 0.45 &  15.2 & 0.13 &  2.36 & 0.02 &      &      &      &      &      &       \\
ATLAS $W,Z$ 7~TeV CC
& \cite{ATLAS:2016nqi}
&  9.0
&  94.4 & 0.50 &  21.9 & 0.19 &  8.63 & 0.09 & 4.15 & 0.05 & 2.12 & 0.02 & 0.50 & 0.01  \\
ATLAS $W,Z$ 7~TeV CF
& \cite{ATLAS:2016nqi}
&  2.8
&  69.7 & 0.49 &  9.13 & 0.10 &       &      &      &      &      &      &      &       \\
ATLAS LM DY 2D 8~TeV
& \cite{ATLAS:2017rue}
&  1.1
&  15.1 & 0.09 &       &      &       &      &      &      &      &      &      &       \\
ATLAS HM DY 2D 8~TeV
& \cite{ATLAS:2016gic}
&  2.6
&  79.6 & 0.40 &  8.42 & 0.08 &       &      &      &      &      &      &      &       \\
ATLAS $\sigma_{W,Z}^{\rm tot}$ 13~TeV
& \cite{ATLAS:2016fij}
&  5.0
&  57.6 & 0.60 &  11.1 & 0.18 &  3.72 & 0.07 & 1.15 & 0.02 &      &      &      &       \\
ATLAS $W^+$+jet 8~TeV
& \cite{ATLAS:2017irc}
&  4.0
&  78.7 & 0.52 &  12.2 & 0.16 &  2.55 & 0.05 & 0.03 &      &      &      &      &       \\
ATLAS $W^-$+jet 8~TeV
& \cite{ATLAS:2017irc}
&  5.7
&  82.0 & 0.54 &  14.9 & 0.18 &  4.74 & 0.07 & 1.78 & 0.03 & 0.51 & 0.01 &      &       \\
ATLAS $Z$ $p_T,m_{\ell\ell}$ 8~TeV
& \cite{ATLAS:2015iiu}
&  3.3
&  71.6 & 0.44 &  7.99 & 0.09 &       &      &      &      &      &      &      &       \\
ATLAS $Z$ $p_T,y_Z$ 8~TeV
& \cite{ATLAS:2015iiu}
&  8.8
&  83.1 & 0.43 &  11.6 & 0.11 &       &      &      &      &      &      &      &       \\
ATLAS $t\bar{t}~\ell$+jets $y_t$ 8~TeV
& \cite{ATLAS:2015lsn}
&  1.6
&  37.3 & 0.34 &       &      &       &      &      &      &      &      &      &       \\
ATLAS $t\bar{t}~\ell$+jets $y_{t\bar t}$ 8~TeV
& \cite{ATLAS:2015lsn}
&  2.2
&  52.5 & 0.41 &  2.31 & 0.03 &       &      &      &      &      &      &      &       \\
ATLAS $t\bar{t}~2\ell$ $y_{t\bar t}$ 8~TeV
& \cite{ATLAS:2016pal}
&  1.9
&  31.5 & 0.28 &       &      &       &      &      &      &      &      &      &       \\
ATLAS jets R=0.6 8~TeV
& \cite{ATLAS:2017kux}
&  5.5
&  93.6 & 0.48 &  20.6 & 0.17 &  7.21 & 0.07 & 2.61 & 0.03 & 0.53 & 0.01 &      &       \\
ATLAS dijets R=0.6 7~TeV
& \cite{ATLAS:2013jmu}
&  10
&  95.7 & 0.51 &  22.3 & 0.19 &  9.08 & 0.09 & 4.54 & 0.05 & 2.53 & 0.03 & 0.80 & 0.01  \\
ATLAS $\gamma$ 13 TeV
& \cite{ATLAS:2017nah}
&  1.3
&  38.2 & 0.14 &       &      &       &      &      &      &      &      &      &       \\
ATLAS single~$t$ $dy_t$ 7~TeV
& \cite{ATLAS:2014sxe}
&  1.3
&  20.8 & 0.22 &       &      &       &      &      &      &      &      &      &       \\
ATLAS single~$t$ $y_{\bar t}$ 7~TeV
& \cite{ATLAS:2014sxe}
&  1.4
&  29.7 & 0.25 &       &      &       &      &      &      &      &      &      &       \\
ATLAS single~$t$ $dy_t$ 8~TeV
& \cite{ATLAS:2017rso}
&  1.2
&  13.5 & 0.12 &       &      &       &      &      &      &      &      &      &       \\
ATLAS single~$t$ $y_{\bar t}$ 8~TeV
& \cite{ATLAS:2017rso}
&  1.2
&  15.6 & 0.17 &       &      &       &      &      &      &      &      &      &       \\
\midrule
CMS $W$ $e$ asy. 7~TeV
& \cite{CMS:2012ivw}
&  1.0
&  9.17 & 0.03 &       &      &       &      &      &      &      &      &      &       \\
CMS $W$ $\mu$ asy. 7~TeV
& \cite{CMS:2013pzl}
&  1.2
&  30.3 & 0.12 &       &      &       &      &      &      &      &      &      &       \\
CMS DY 2D 7~TeV
& \cite{CMS:2013zfg}
&  8.8
&  85.2 & 0.52 &  17.0 & 0.18 &  5.99 & 0.08 & 2.86 & 0.04 & 1.53 & 0.02 & 0.42 & 0.01  \\
CMS $W$ rapidity 8~TeV
& \cite{CMS:2016qqr}
&  13
&  93.9 & 0.59 &  22.6 & 0.21 &  9.39 & 0.10 & 4.88 & 0.05 & 2.79 & 0.03 & 1.03 & 0.01  \\
CMS $Z$ $p_T$ 8~TeV
& \cite{CMS:2015hyl}
&  9.5
&  87.7 & 0.46 &  15.9 & 0.14 &  3.09 & 0.03 &      &      &      &      &      &       \\
CMS dijets 7~TeV
& \cite{CMS:2012ftr}
&  4.7
&  88.8 & 0.48 &  18.2 & 0.15 &  5.51 & 0.05 & 1.22 & 0.01 &      &      &      &       \\
CMS jets 8~TeV
& \cite{CMS:2016lna}
&  6.3
&  92.4 & 0.53 &  20.2 & 0.19 &  7.14 & 0.08 & 2.78 & 0.03 & 0.93 & 0.01 &      &       \\
CMS $t\bar{t}~\ell$+jets 8~TeV
& \cite{CMS:2015rld}
&  1.6
&  41.9 & 0.26 &       &      &       &      &      &      &      &      &      &       \\
CMS $t\bar{t}$ 2D $2\ell$ 8~TeV
& \cite{CMS:2017iqf}
&  1.9
&  58.3 & 0.33 &  5.84 & 0.06 &       &      &      &      &      &      &      &       \\
CMS $t\bar{t}~2\ell$ 13~TeV
& \cite{CMS:2018adi}
&  5.2
&  76.6 & 0.48 &  11.5 & 0.13 &  2.11 & 0.03 & 0.51 & 0.01 & 0.05 &      &      &       \\
CMS $t\bar{t}~\ell$+jet 13~TeV
& \cite{CMS:2018htd}
&  7.5
&  83.3 & 0.51 &  17.7 & 0.17 &  6.36 & 0.07 & 2.67 & 0.03 & 1.06 & 0.01 & 0.07 &       \\
\midrule
LHCb $Z\to ee$ 7~TeV
& \cite{LHCb:2012gii}
&  1.4
&  55.1 & 0.34 &       &      &       &      &      &      &      &      &      &       \\
LHCb $W,Z \to \mu$ 7~TeV
& \cite{LHCb:2015okr}
&  2.9
&  66.6 & 0.40 &  5.22 & 0.08 &       &      &      &      &      &      &      &       \\
LHCb $Z\to ee$ 8~TeV
& \cite{LHCb:2015kwa}
&  1.4
&  45.3 & 0.20 &       &      &       &      &      &      &      &      &      &       \\
LHCb $W,Z\to \mu$ 8~TeV
& \cite{LHCb:2015mad}
&  2.5
&  69.8 & 0.43 &  5.28 & 0.07 &       &      &      &      &      &      &      &       \\
LHCb $Z\to ee$ 13~TeV
& \cite{LHCb:2016fbk}
&  2.4
&  54.9 & 0.26 &       &      &       &      &      &      &      &      &      &       \\
LHCb $Z\to \mu\mu$ 13~TeV
& \cite{LHCb:2016fbk}
&  1.6
&  74.9 & 0.39 &  5.69 & 0.06 &       &      &      &      &      &      &      &       \\
\bottomrule
\end{tabularx}

%% file: tables/tab-chi2.tex
\begin{tabularx}{\textwidth}{Xrccccccc}
\toprule
& & \multicolumn{7}{c}{$\chi^2/N_{\rm dat}$}\\
Data set
& $N_{\rm dat}$
& NNPDF4.0
& $\delta^{-1}=1$
& $\delta^{-1}=2$
& $\delta^{-1}=3$
& $\delta^{-1}=4$
& $\delta^{-1}=5$
& $\delta^{-1}=7$ \\
\midrule
Deep-inelastic scattering
&  3089
& 1.12 & 0.64 & 1.02 & 1.09 & 1.11 & 1.12 & 1.12 \\
Fixed-target Drell-Yan
&   195
& 0.98 & 0.48 & 0.90 & 0.96 & 0.97 & 0.97 & 0.99 \\
Tevatron Drell-Yan
&    65
& 1.11 & 0.48 & 0.71 & 0.85 & 0.93 & 1.02 & 1.10 \\
\midrule
ATLAS total
&   679
& 1.24 & 0.50 & 0.84 & 0.97 & 1.04 & 1.10 & 1.19 \\
\ \ \ LM DY 7~TeV
&     6
& 0.88 & 0.22 & 0.60 & 0.85 & 0.88 & 0.88 & 0.88 \\
\ \ \ HM DY 7~TeV
&     5
& 1.69 & 0.44 & 1.25 & 1.64 & 1.69 & 1.69 & 1.69 \\
\ \ \ $W,Z$ 7~TeV
&    30
& 0.98 & 0.24 & 0.65 & 0.96 & 1.01 & 1.01 & 1.00 \\
\ \ \ $W,Z$ 7~TeV CC
&    46
& 1.92 & 0.31 & 0.74 & 0.94 & 1.21 & 1.47 & 1.76 \\
\ \ \ $W,Z$ 7~TeV FC
&    15
& 1.03 & 0.48 & 0.96 & 1.04 & 1.04 & 1.04 & 1.04 \\
\ \ \ LM DY 2D 8~TeV
&    48
& 1.11 & 0.61 & 1.09 & 1.12 & 1.11 & 1.11 & 1.12 \\
\ \ \ HM DY 2D 8~TeV
&    60
& 1.22 & 1.18 & 1.21 & 1.22 & 1.22 & 1.22 & 1.22 \\
\ \ \ $\sigma_{W,Z}^{\rm tot}$ 13~TeV
&     3
& 0.77 & 0.30 & 0.39 & 0.46 & 0.62 & 0.79 & 0.77 \\
\ \ \ $W^+$+jet 8~TeV
&    15
& 0.79 & 0.45 & 0.58 & 0.70 & 0.79 & 0.79 & 0.79 \\
\ \ \ $W^-$+jet 8~TeV
&    15
& 1.49 & 0.79 & 1.15 & 1.29 & 1.38 & 1.44 & 1.49 \\
\ \ \ $Z$ $p_T,m_{\ell\ell}$ 8~TeV
&    44
& 0.90 & 0.43 & 0.86 & 0.90 & 0.91 & 0.90 & 0.90 \\
\ \ \ $Z$ $p_T,y_Z$ 8~TeV
&    48
& 0.90 & 0.22 & 0.65 & 0.91 & 0.90 & 0.90 & 0.90 \\
\ \ \ $\sigma_{t\bar{t}}$ 7, 8, 13~TeV
&     3
& 1.64 & 1.47 & 1.70 & 1.74 & 1.79 & 1.77 & 1.70 \\
\ \ \ $t\bar{t}~\ell$+jets $y_t$ 8~TeV
&     4
& 3.28 & 1.42 & 2.45 & 2.92 & 2.98 & 3.01 & 3.09 \\
\ \ \ $t\bar{t}~\ell$+jets $y_{t\bar t}$ 8~TeV
&     4
& 3.83 & 1.00 & 2.62 & 3.32 & 3.40 & 3.46 & 3.52 \\
\ \ \ $t\bar{t}~2\ell$ $y_{t\bar t}$ 8~TeV
&     5
& 1.62 & 0.65 & 1.46 & 1.55 & 1.57 & 1.59 & 1.59 \\
\ \ \ jets R=0.6 8~TeV
&   171
& 0.68 & 0.38 & 0.60 & 0.68 & 0.69 & 0.69 & 0.69 \\
\ \ \ dijets R=0.6 7~TeV
&    90
& 2.14 & 0.23 & 0.59 & 0.85 & 1.10 & 1.38 & 1.87 \\
\ \ \ $\gamma$ 13 TeV
&    53
& 0.76 & 0.58 & 0.72 & 0.76 & 0.77 & 0.76 & 0.76 \\
\ \ \ single $t$ $R_t$~7, 13~TeV
&     2
& 0.28 & 0.24 & 0.27 & 0.28 & 0.28 & 0.28 & 0.28 \\
\ \ \ single~$t$ $dy_t$ 7~TeV
&     3
& 0.96 & 0.94 & 0.98 & 0.97 & 0.96 & 0.96 & 0.96 \\
\ \ \ single~$t$ $y_{\bar t}$ 7~TeV
&     3
& 0.06 & 0.03 & 0.06 & 0.06 & 0.06 & 0.06 & 0.06 \\
\ \ \ single~$t$ $dy_t$ 8~TeV
&     3
& 0.25 & 0.20 & 0.23 & 0.24 & 0.24 & 0.24 & 0.24 \\
\ \ \ single~$t$ $y_{\bar t}$ 8~TeV
&     3
& 0.19 & 0.17 & 0.19 & 0.19 & 0.19 & 0.19 & 0.19 \\
\midrule
CMS total
&   474
& 1.31 & 0.39 & 0.83 & 1.08 & 1.21 & 1.26 & 1.28 \\
\ \ \ $W$ $e$ asy. 7~TeV
&    11
& 0.84 & 0.71 & 0.79 & 0.79 & 0.79 & 0.79 & 0.81 \\
\ \ \ $W$ $\mu$ asy. 7~TeV
&    11
& 1.70 & 1.34 & 1.76 & 1.76 & 1.76 & 1.76 & 1.75 \\
\ \ \ DY 2D 7~TeV
&   110
& 1.36 & 0.43 & 1.05 & 1.27 & 1.33 & 1.35 & 1.36 \\
\ \ \ $W$ rapidity 8~TeV
&    22
& 1.33 & 0.13 & 0.16 & 0.22 & 0.33 & 0.46 & 0.80 \\
\ \ \ $Z$ $p_T$ 8~TeV
&    28
& 1.40 & 0.21 & 0.75 & 1.25 & 1.40 & 1.40 & 1.40 \\
\ \ \ dijets 7~TeV
&    54
& 1.79 & 0.63 & 1.40 & 1.73 & 1.84 & 1.84 & 1.82 \\
\ \ \ jets 8~TeV
&   185
& 1.19 & 0.24 & 0.53 & 0.84 & 1.07 & 1.15 & 1.18 \\
\ \ \ $\sigma_{t\bar{t}}$ 5, 8, 7, 13~TeV
&     4
& 0.45 & 0.39 & 0.49 & 0.51 & 0.53 & 0.52 & 0.48 \\
\ \ \ $t\bar{t}~\ell$+jets 8~TeV
&     9
& 1.25 & 0.67 & 1.13 & 1.22 & 1.23 & 1.23 & 1.23 \\
\ \ \ $t\bar{t}$ 2D $2\ell$ 8~TeV
&    16
& 1.01 & 0.41 & 0.86 & 1.07 & 1.07 & 1.06 & 1.04 \\
\ \ \ $t\bar{t}~2\ell$ 13~TeV
&    10
& 0.52 & 0.09 & 0.26 & 0.37 & 0.43 & 0.50 & 0.51 \\
\ \ \ $t\bar{t}~\ell$+jet 13~TeV
&    11
& 0.64 & 0.04 & 0.15 & 0.33 & 0.42 & 0.51 & 0.59 \\
\ \ \ single $t$ $R_t$ 7, 8, 13~TeV
&     3
& 0.42 & 0.38 & 0.41 & 0.42 & 0.42 & 0.42 & 0.42 \\
\midrule
LHCb total
&   116
& 1.55 & 0.73 & 1.41 & 1.53 & 1.56 & 1.55 & 1.55 \\
\ \ \ $Z\to ee$ 7~TeV
&     9
& 1.64 & 0.88 & 1.55 & 1.62 & 1.63 & 1.63 & 1.65 \\
\ \ \  $W,Z \to \mu$ 7~TeV
&    29
& 1.94 & 0.66 & 1.61 & 1.93 & 1.97 & 1.97 & 1.96 \\
\ \ \ $Z\to ee$ 8~TeV
&    17
& 1.34 & 0.88 & 1.30 & 1.31 & 1.33 & 1.33 & 1.33 \\
\ \ \ $W,Z\to \mu$ 8~TeV
&    30
& 1.42 & 0.73 & 1.31 & 1.41 & 1.45 & 1.43 & 1.43 \\
\ \ \  $Z\to ee$ 13 TeV
&    15
& 1.73 & 0.95 & 1.71 & 1.72 & 1.73 & 1.73 & 1.72 \\
\ \ \  $Z\to \mu\mu$ 13 TeV
&    16
& 1.00 & 0.31 & 0.91 & 0.98 & 0.99 & 0.99 & 0.99 \\
\midrule
Total
&  4618
& 1.16 & 0.58 & 0.97 & 1.07 & 1.11 & 1.13 & 1.15 \\
\bottomrule
\end{tabularx}

%% file: tables/tab-chi2_models.tex
\begin{tabularx}{\textwidth}{Xrcccccc}
\toprule
& & \multicolumn{5}{c}{$\chi^2/N_{\rm dat}$}\\
Data set
& $N_{\rm dat}$
& NNPDF4.0
& STRONG
& WEAK
& ATLAS
& ATLAS+WEAK
& $\delta^{-1}=4$\\
\midrule
Deep-inelastic scattering
&  3089
& 1.12 & 1.12 & 1.12 & 1.12 & 1.12 & 1.11 \\
Fixed-target Drell-Yan
&   195
& 0.98 & 1.00 & 0.99 & 0.99 & 0.99 & 0.97 \\
Tevatron Drell-Yan
&    65
& 1.11 & 1.10 & 1.09 & 1.09 & 1.10 & 0.93 \\
\midrule
ATLAS total
&   679
& 1.24 & 1.24 & 1.24 & 1.23 & 1.24 & 1.04 \\
\ \ \ $W^+$+jet 8~TeV
&    15
& 0.79 & 0.78 & 0.79 & 0.79 & 0.79 & 0.79 \\
\ \ \ $W^-$+jet 8~TeV
&    15
& 1.49 & 1.49 & 1.49 & 1.49 & 1.50 & 1.38 \\
\ \ \ $t\bar{t}~\ell$+jets $y_t$ 8~TeV
&     4
& 3.28 & 3.14 & 3.06 & 3.04 & 3.16 & 2.98 \\
\ \ \ $t\bar{t}~\ell$+jets $y_{t\bar t}$ 8~TeV
&     4
& 3.83 & 3.58 & 3.57 & 3.54 & 3.65 & 3.40 \\
\ \ \ jets R=0.6 8~TeV
&   171
& 0.68 & 0.69 & 0.68 & 0.68 & 0.68 & 0.69 \\
\ \ \ dijets R=0.6 7~TeV
&    90
& 2.14 & 2.16 & 2.15 & 2.14 & 2.15 & 1.10 \\
\midrule
CMS total
&   474
& 1.31 & 1.31 & 1.31 & 1.31 & 1.30 & 1.21 \\
\midrule
LHCb total
&   116
& 1.55 & 1.56 & 1.54 & 1.55 & 1.55 & 1.56 \\
\midrule
Total
&  4618
& 1.16 & 1.16 & 1.16 & 1.16 & 1.16 & 1.11 \\
\bottomrule
\end{tabularx}

%% file: sec-conclusions.tex
\section{Conclusions}
\label{sec:conclusions}

In this paper we have shown how an (even slightly) inaccurate determination of 
bin-by-bin correlations in the uncertainties of experimental measurements may
make the $\chi^2$ statistic fluctuate substantially, by more than one standard
deviation. This problem is particularly relevant when dealing with
high-precision measurements, in which the largest fraction of the uncertainty is
correlated. This is the case for current and future LHC measurements that are
routinely confronted with theoretical predictions by means of statistical
inference. Because the $\chi^2$ is routinely utilised as a figure of merit in
these analyses, instabilities in its computation can make the interpretation of
the results unreliable.

We have formulated the problem rigorously, by deriving a stability criterion
for the acceptable fluctuations of the uncertainties on data correlations. The
criterion ensures that the expectation value of the $\chi^2$ does not
overestimate its true value by an amount larger than its statistical
fluctuation. To this aim, the criterion defines a bound on the singular values
of the correlated part of the matrix of uncertainties. Building upon this
criterion, we have then devised a regularisation procedure, whereby
instabilities in the correlations of experimental uncertainties are removed
with minimal information and without loss of generality. The idea is to clip
the singular values of the correlated part of the matrix of uncertainties to a
constant $\delta$, whenever these are smaller than that, while leaving the
rest of the singular vectors unchanged. This way, directions that do not
contribute to instability are not affected and the alteration to the original
matrix is minimal.

The key assumptions underlying the regularisation procedure are that
correlations of experimental uncertainties across data points are determined
much less precisely than the uncertainties on each data point, and that the
prevalent source of inaccuracy on correlations concentrates on a subset of data
points and originates from a small number of correlated uncertainties. The
regularisation procedure leads to a covariance matrix that is more stable than
the original one, when used to compute the $\chi^2$, is compatible with it
within the precision with which it is determined, and does not lead to a
reduction of the total uncertainty.

We have demonstrated how the regularisation procedure works in a toy model,
and in a particular problem relevant to LHC precision physics that relies on
the evaluation of the $\chi^2$ as a figure of merit: PDF determination.
Specifically, we have considered the NNPDF4.0
determination~\cite{NNPDF:2021njg},
which is based on the widest data set to date. We have shown how the
regularisation procedure can be utilised as a diagnosis tool to characterise
the data set, in particular to single out those measurements for which an
inaccurate estimation of experimental correlations may significantly affect
their $\chi^2$. We have also studied how PDFs change if the nominal data set is
replaced by a suitably regularised data set, and how these changes depend on the
regularisation parameter $\delta$. To this purpose, we have repeated the
NNPDF4.0 baseline fit, now utilising a data set regularised with
$\delta^{-1}=1,2,3,4,5,7$.

We have found that the $\chi^2$ of some LHC data sets can be indeed
significantly affected by inaccuracies in the determination of the
correlations of their uncertainties. These inaccuracies can be reasonably
regularised by choosing $\delta^{-1}=4$. This value sets the precision
with which uncertainties and correlations are known to less than $5\%$ and
$0.05$, respectively. We have demonstrated that, by regularising the NNPDF4.0
data set with $\delta^{-1}=4$, the global $\chi^2$ is smaller than that of the
baseline NNPDF4.0 determination by about $2.4\sigma$. This means that it is only
$5.3\sigma$ away from the unity expectation (instead of $7.7\sigma$ in the
baseline NNPDF4.0 determination). At the same time, PDFs remain unaltered.
These results highlight the fact that the nominal $\chi^2$ determined
in~\cite{NNPDF:2021njg} is likely to be spuriously inflated by inaccuracies in
the estimation of the experimental correlations in the LHC data.

Finally, we have studied how the regularisation procedure performs in
comparison to correlation models provided with the measurements in the few
cases in which these are available. We have found that our regularisation
procedure captures the same qualitative shifts on PDF central values as
experimental correlation models, once one or the others are used to determine
the PDFs. Whether this is a coincidental feature, limited to the correlation
models considered, or a more general one, could only be investigated if
additional correlation models become available to be tested.
On the other hand, using correlation models does not lead to a better
$\chi^2$ (or to a decrease in the condition number of the corresponding
experimental covariance matrix). We therefore conclude that the available
correlation models are not enough to lead to a stable $\chi^2$. In light of
these considerations, we find that our regularisation procedure can
possibly be used as a useful diagnosis tool to inform and validate correlation
models not only at the level of PDF fits, but also at the level of the
corresponding experimental analyses.

\vspace{0.5cm}

\noindent Our regularisation procedure is made publicly available as part of
the {\sc NNPDF} software~\cite{NNPDF:2021uiq}. The PDF sets discussed in this
paper are available, in the {\sc LHAPDF} format~\cite{Buckley:2014ana},
from the NNPDF web page~\cite{NNPDF:regularisation}.

%% file: sec-acknowledgements.tex
\section*{Acknowledgements}
\label{sec:acknowledgements}

We thank our colleagues of the NNPDF Collaboration for useful discussions, in
particular Alessandro Candido and Christopher Schwan for a careful reading
of the manuscript.
Z.~K. is supported by the European Research Council under the European Union's
Horizon 2020 research and innovation Programme (grant agreement n.950246).
E.~R.~N. is supported by the U.K. Science and Technology Facility Council
(STFC) grant ST/P000630/1. M.~W. was supported by the STFC grant ST/R504737/1.

%% file: app-glossary.tex
\section{Glossary}
\label{app:glossary}

In this Appendix we recall the linear algebra definitions and the notation 
that are used throughout the paper. Let $A$ be a $N \times M$ real matrix,
$A \in \mathbb{R}^{N\times M}$, such as the error matrix introduced in
Sect.~\ref{sec:stability}. Then the following statements hold.

\begin{description}

\item[Singular value decomposition.] The matrix $A$ admits a singular value
  decomposition of the form
  \begin{equation}
    A = USV^t
    \,,
    \label{eq:svd}
  \end{equation} 
  where $U$ and $V$ are $N \times N$ and $M \times M$ orthogonal matrices,
  respectively, and $S$ is a $N \times M$ diagonal matrix with non negative real
  numbers in the principal diagonal. The diagonal entries of $S$,
  $s_i,\ i = 1,\dots,\min(N,M)$, are the singular values of $A$, ordered by
  decreasing size, with $s_{\max} = s_1$ being the largest singular value.

\item[Rank.] The rank of the matrix $A$, denoted as $\rank(A)$, is the number
  of strictly positive singular values. The matrix $A$ if full rank if its rank
  is $\min(N,M)$.

\item[Pseudoinverse.] The pseudoinverse of the matrix $A$ is
  \begin{equation}
    A^+ = VS^+U^t
    \,,
    \label{eq:pseudoinversedef}
  \end{equation}
  where $S^+$ is a $M \times N$ diagonal matrix with entries
  \begin{equation}
    S_{ii}^{+}=\begin{cases}
    1/s_{i} & s_{i}\neq0\\
    0 & s_{i}=0
    \end{cases}
    \,,
    \ i = 1,\dots, \min(N, M)
    \,.
    \label{eq:entries}
  \end{equation}

\item[Right inverse.] The right inverse of the matrix $A$ (which we denote
  with the same notation used for the pseudoinverse) is the $M\times N$
  matrix $A^+$ such that
  \begin{equation}
    AA^+ = I_{N\times N}
    \,.
    \label{eq:right_inverse}
  \end{equation}
  
\item[Euclidean norm.] The Euclidean (or vector-induced $L^2$) norm of the
  matrix $A$ is
  \begin{equation}
    \norm{A}_2 =  \max_{
      \left\{
      \vb{x}\in \mathbb{R}^M  : \norm{\vb{x}} = 1
      \right\}
    } 
    \norm{A\vb{x}
    } 
    = s_{\max}
    \,.
    \label{eq:l2def}
  \end{equation}

\item[Frobenius norm.] The Frobenius norm of the matrix $A$ is
  \begin{equation}
    \norm{A}_F = \sqrt{\sum_i\sum_j A_{ij}^2} = \sqrt{\trace(AA^t)} = \sqrt{\sum_i^{\rank A}s_i^2}
    \,.
    \label{eq:frobeniusdef}
  \end{equation}
  In Sects.~\ref{sec:stability}-\ref{sec:regularisation} we always indicate
  with a subindex whether we are referring to the Euclidean or the Frobenius
  norm of the matrix $A$. By noting that $s_i \leq s_{\max}$, it follows from
  Eqs.~\eqref{eq:l2def} and~\eqref{eq:frobeniusdef} that
  \begin{equation}
    \norm{A}_2 \leq \norm{A}_F \leq \sqrt{\rank(A)} \norm{A}_2    
    \,.
    \label{eq:normorder}
  \end{equation}

\end{description}

%% file: app-boundproof.tex
\section{Proof of Eq.~\eqref{eq:norminequality}}
\label{app:boundproof}

In this Appendix, we provide a proof of Eq.~\eqref{eq:norminequality}.
Let $X$ and $Y$ be arbitrary matrices of dimensions $M\times N $
and $N\times M$, respecctively. We first show that
\begin{equation}
  \norm{XY}_F \leq \norm{X}_F\norm{Y}_2
  \,.
\end{equation}

We write the respective singular value decompositions, $X = U_XS_XV^t_X$ and
$Y = U_Y S_Y V^t_Y$, and define
\begin{equation}
  W = V^t_X U_Y 
  \,,
\end{equation}
which is orthogonal because both factors are. Then
\begin{equation}
  \norm{XY}_F^2 = \norm{S_X W S_Y}_F^2    
  \,.
\end{equation}
Hence
\begin{equation}
  \norm{XY}_F^2  = \sum_{i}\sum_j \left(S_{X(ii)} W_{(ij)} S_{Y(jj)}\right)^2   
  \,, 
\end{equation}
where the sums are implied to run over the relevant matrix dimensions $M$ and
$N$. We bind the singular values $S_{Y(jj)}$ by their maximum, the Euclidean
norm of Y. That is
\begin{equation}
  \norm{XY}_F^2
  \leq
  \sum_i \left(S_{X(ii)}^2 \sum_j W_{(ij)}^2 \norm{Y}_2^2 \right)
  \,.
\end{equation}
Because $W$ is orthogonal, it follows that
\begin{equation}
  \norm{XY}_F^2
  \leq
  \left(\sum_i S^2_{X(ii)}\right)\norm{Y}_2^2
  =
  \norm{X}_F^2\norm{Y}_2^2
  \,.
\end{equation}
The converse relation
\begin{equation}
  \norm{XY}_F \leq \norm{X}_2\norm{Y}_F
\end{equation}
can be proven analogously. Hence we obtain the result,
Eq.~\eqref{eq:norminequality},
\begin{equation}
  \norm{XY}_F  \leq \min\left(\norm{X}_F\norm{Y}_2, \norm{X}_2\norm{Y}_F \right)
  \,.
\end{equation}